\documentclass[12pt,preprint]{aastex}
\begin{document}

\title{The Disk and Extraplanar Regions of NGC 55}

\author{T. J. Davidge\altaffilmark{1}\altaffilmark{2}\altaffilmark{3}}

\affil{Canadian Gemini Office, Herzberg Institute of Astrophysics,
\\National Research Council of Canada, 5071 West Saanich Road,
\\Victoria, B.C. Canada V9E 2E7}

\email{tim.davidge@nrc.ca}

\altaffiltext{1}{Based on observations obtained at the
Gemini Observatory, which is operated by the Association of Universities
for Research in Astronomy, Inc., under a co-operative agreement with the
NSF on behalf of the Gemini partnership: the National Science Foundation
(United States), the Particle Physics and Astronomy Research Council
(United Kingdom), the National Research Council of Canada (Canada),
CONICYT (Chile), the Australian Research Council (Australia), CNPq (Brazil),
and CONICET (Argentina).} 

\altaffiltext{2}{Visiting Astronomer, Canada-France-Hawaii
Telescope, which is operated by the National Research Council of Canada,
the Centre National de la Recherche Scientifique, and the University of
Hawaii.}

\altaffiltext{3}{This publication makes use of data products from the Two Micron All 
Sky Survey, which is a joint project of the University of Massachusetts and the 
Infrared Processing and Analysis Center/California Institute of Technology, funded 
by the National Aeronautics and Space Administration and the National Science 
Foundation.}

\begin{abstract}

	The stellar content of the nearby 
SB(s)m galaxy NGC 55 is investigated using images obtained with the Gemini 
South and Canada-France-Hawaii telescopes. The $(K, H-K)$ 
and $(K, J-K)$ color-magnitude diagrams (CMDs) 
of stars near the plane of the disk reveal signatures of large scale 
star formation during recent and intermediate epochs in the form of red 
supergiants (RSGs) with M$_K \sim -11.5$, and an asymptotic giant branch 
(AGB) that peaks near M$_K = -10$. Comparisons with stellar evolution models suggest 
that the brightest RSGs have an age near 8 Myr. 
A well-defined plume, which stellar evolution models suggest 
contains stars with masses near the RSG -- AGB transition, 
is detected in CMDs constructed both from infrared and visible 
wavelength observations. It is concluded that star formation in the thin disk 
of NGC 55 has ocurred at a significant rate for at least the past 0.1 -- 0.2 
Gyr, and this is consistent with other indicators. 
The near-infrared spectral-energy distribution (SED) of the integrated light 
near the center of the galaxy is consistent with that in other Magellanic 
irregular galaxies, indicating that the star-forming history of NGC 55, when 
averaged over timescales of 0.1 -- 1 Gyr, has likely not been peculiar when 
compared with other late-type systems. Evidence is also presented that the 
disk contains a large population of old (log(t$_{yr}$) $\sim 10$) stars, 
and it is argued that a stable disk has been in place in NGC 55 for a significant 
fraction of the age of the Universe. 

	At projected distances in excess of 2 kpc off of the disk plane the brightest 
AGB stars have ages $10^{+3}_{-2}$ Gyr. Thus, despite indications that dust and 
gas are present in the envelope surrounding the NGC 55 disk, the AGB content suggests 
that recently formed stars do not occur in large numbers in the extraplanar region. 
The $(r'-i')$ colors of the RGB in the extraplanar region are 
consistent with [Fe/H] between --2.2 and --0.7, with the majority of stars having 
[Fe/H] $> -1.2$, and the mean metallicity inferred from the RGB color does not 
change with distance above the disk plane. 
Thus, the stellar component in the extraplanar envelope is 
well mixed, at least in terms of metallicity. The mean metallicity of RGB stars 
is in excellent agreement with that measured in the extraplanar HII regions 
EHR 1 and 2, suggesting that the age-metallicity relation in this part of NGC 55 
has been flat for at least a few Gyr. Finally, the RGB-tip occurs near $i' = 23.1$ 
in the extraplanar region, and a distance modulus of 26.5 is computed from 
this feature. 

\end{abstract}


\section{INTRODUCTION}

	A current prevailing view is that large galaxies formed from 
the hierarchal accretion of smaller systems. In this picture, the 
first proto-galactic fragments were likely disks that had masses significantly 
smaller than the Milky-Way (Klypin et al. 1999), and obtained 
angular momentum from the torquing action of nearby mass 
concentrations. These early disks were consumed quickly 
in major mergers that kinematically heated the 
progenitors and induced large-scale star formation. 

	The merger activity that dominated during early epochs could have prevented the 
formation of stable, long-lived disks (e.g. Weil, Eke, \& Efstathiou 1998), and the 
epoch when stable disks appear in large numbers is of great importance for 
understanding the evolution of disk-dominated galaxies. 
Efforts to measure the age of the Galactic disk have yielded a range 
of results (e.g. discussion in Hansen 1999), and so studies of other galaxies are of 
great interest. Systems with disk-like light distributions are seen in large 
numbers at $z > 2$ (e.g. Marleau \& Simard 1998; Labbe et al. 2003), 
suggesting that they may be long-lived structures even at high redshifts. 
Among nearby galaxies, an early epoch for the formation of stable disks 
is supported by the presence of metal-poor, and hence presumably old, globular clusters 
that appear to be associated with the thin disk of M31 (Morrison et al. 
2004). The presence of such a cluster system in M31 may indicate that 
disks are relatively hardy entities, given the evidence 
for tidal interactions between M31 and its companions (Ferguson et al. 2002), 
which may have caused large-scale mixing (Brown et al. 2003; Bellazzini et al. 2003). 

	While long-lived disks may not form early-on, the extraplanar 
regions of spiral galaxies may contain stars with ages that 
span the time from the collapse of the first 
proto-galactic fragments to the epoch of disk formation and beyond. 
Efforts to simulate the formation of the Milky-Way predict that the 
halo started to form during early merging events that 
were tied to the creation of the bulge, and preceded the 
formation of the disk (e.g. Nakasato \& Nomoto 2003; Bekki \& Chiba 2001). 
A number of processes, such as the continued accretion of protogalactic 
material (Bekki \& Chiba 2001), the disruption of 
satellites and the heating of the disk by 
tidal interactions (e.g. Sellwood, Nelson, \& 
Tremaine 1998), as well as the ejection of gas outside of the disk plane 
by star forming events, which may cool so that stars can form 
(e.g. Savage et al. 2003), have the potential to add stars with a range of 
ages (and metallicities) to extraplanar environments.

	Edge-on disk galaxies are important laboratories 
for investigating the evolutionary connections between halos, disks, and 
bulges. If halo and bulge formation are linked, then galaxies 
with very small bulges should have correspondingly small halos if compared with 
galaxies of similar size but earlier morphological type. On the other 
hand, if disk and halo formation are coupled then the relative sizes of halos 
and bulges should not be connected.

	Dalcanton \& Bernstein (2002) investigated the integrated photometric 
properties of undisturbed bulgeless edge-on disks, and found that they are 
embedded in diffuse flattened envelopes that show remarkable galaxy-to-galaxy 
similarities. The isophotes tend to become 
less flattened with increasing distance off of the disk, and 
the integrated photometric properties of the envelopes are 
consistent with an old population having an age in the range 6 -- 8 Gyr. While 
there is a large range in the metallicities estimated for the envelopes, there are 
roughly equal numbers of systems with positive and negative metallicity 
gradients off the disk, and the mean trend is for metallicity not to 
change with height above the disk (e.g. Figure 10 of Dalcanton \& Bernstein 
2002). Given (1) the flattened nature of these envelopes and 
(2) that the target galaxies are bulgeless, Dalcanton \& Bernstein (2002) 
associate the envelope with the disk, and argue that it is an artifact of 
merger activity that kinematically heated disk material. 

	The nearby collection of galaxies in Sculptor is one of the closest 
ensembles of galaxies outside of the Local Group. The galaxies that are traditionally 
associated with this group extend over at least 1.5 Mpc along the line of 
sight, and may not form a gravitationally bound system (e.g. Karachentsev et 
al. 2003). The brightest members are late-type galaxies, and so the Sculptor group is 
an important laboratory for studying disk-dominated systems. 

	One of the closest members of the Sculptor group is NGC 55. This galaxy is 
inclined at an angle near 81$^o$ (Kiszkurno-Koziej 1988), and so is favourably 
oriented to study the extraplanar region. However, the near edge-on orientation also 
complicates efforts to study the morphology of NGC 55. Sandage \& Tammann (1987) 
classify NGC 55 as an Sc galaxy. However, de Vaucouleurs (1961) concludes that 
NGC 55 is structurally similar to the LMC and that the main light concentration 
at visible wavelengths, which is offset from the geometric center of the 
galaxy, is a bar seen end-on. The SBm classification assigned 
by de Vaucouleurs (1961) is consistent with the HI properties of the galaxy, 
which were investigated by Hummel, Dettmar, \& Wielebinski (1986), and who conclude 
that 20\% of the mass is in the bar, while 80\% is distributed throughout 
the disk. van den Bergh (2004, private communication) classifies NGC 55 as a 
dwarf irregular, and notes that a classification as early as Sc is unlikely given 
the absence of an obvious nucleus. For the current study it is assumed that 
NGC 55 is a Magellanic irregular.

	Only a handful of studies have investigated the resolved stellar 
content of NGC 55. In a pioneering study 
Graham (1982) resolved red giant branch (RGB) stars off of the NGC 55 disk 
plane, demonstrating that a rich extraplanar population is 
present. Graham (1982) computed an upper limit of 26.4 to the distance modulus 
based on the peak red giant brightness. Pritchet et al. (1987) investigated the 
bright red stellar content near one end of the galaxy, and found that the 
luminosity function (LF) of asymptotic giant branch (AGB) stars 
is similar to that in other late-type galaxies. A 
population of C stars was identified, and Pritchet et al. (1987) computed a 
distance modulus of $25.7 \pm 0.1$ based on the mean $I-$band magnitude 
of these objects. Davidge (1998) used infrared images to investigate the bright 
red stellar content in two NGC 55 disk fields. A prominent AGB sequence was 
detected, and the peak AGB brightness was found to be consistent with an age $< 0.1$ 
Gyr. A population of red supergiants (RSGs) with $K = 15$ was also found, from which 
Davidge (1998) computed a distance modulus of $\mu = 26.0$.

	There is active star formation throughout much of the disk of 
NGC 55, and there are large quantities of gas off of the disk plane. Ferguson, Wyse, \& 
Gallagher (1996) and Otte \& Dettmar (1999) found shell structures and 
chimneys that are the signatures of supernovae explosions and 
stellar winds. Some of this gas has evidently cooled sufficiently to form stars, as 
T\"{u}llmann et al. (2003) find two compact HII regions that are 2 -- 3 arcmin 
($\sim 1 - 1.5$ kpc) off of the disk. If star formation 
has occured for an extended period of time in this region of NGC 55 
then stars spanning a range of ages should be present in the extraplanar region.

	In the current paper, the stellar contents of the disk and 
extraplanar regions of NGC 55 are investigated with the goal of providing insight 
into the past history of the galaxy. Two sets of observations are employed. One 
dataset, recorded with the CFHTIR imager on the 3.6 meter Canada-France-Hawaii 
Telescope (CFHT), consists of broad and narrow-band near-infrared 
images of a $3 \times 7$ arcmin strip that slices through the center of 
NGC 55 and continues into the extraplanar region. 
The other dataset, obtained with GMOS on the 8 meter Gemini 
South (GS) telescope, consists of deep $r'$ and $i'$ images of a $4.8 
\times 4.8$ arcmin field that samples the extraplanar region and the 
outskirts of the star forming disk.

	The paper is structured as follows. Details of the observations, 
the data reduction procedures, and the photomeric measurements are presented in 
\S 2. The photometric properties of stars and star clusters 
near the disk plane are discussed in \S 3, while 
the photometric properties of extraplanar stars are examined 
in \S 4. A summary and discussion of the results follows in \S 5.
 
\section{OBSERVATIONS \& REDUCTIONS}

\subsection{CFHTIR Data}

	Three adjoining fields that form a north-south strip 
were imaged during the nights of November 21 - 23 2002
with the CFHTIR camera, which was mounted at the f/8 focus of the 
3.6 metre CFHT. The detector in CFHTIR is 
a $1024 \times 1024$ HgCdTe array. Each pixel subtends 0.21 arcsec on a 
side, and so the detector covers a $3.6 \times 3.6$ arcmin$^2$ area.

	A section of the DSS showing the area that was observed with the 
CFHTIR is shown in Figure 1. The entire strip was imaged through $J, H,$ and 
$K'$ filters, with each of the three tiles being observed with a 4 point square 
dither pattern. Additional images of the southernmost tile, which samples 
the star-forming disk, were also recorded through narrow-band CO and 
$K$ continuum filters. The total exposure time was 12 minutes per filter, 
with the exception of the observations through the CO filter, where the total 
exposure time was 24 minutes. The image quality in the processed images is 
FWHM = 1.0 arcsec.

	A series of dark frames were recorded at the end of each night. 
There was no measureable night-to-night variation in the dark count rate during the 
three night observing run, and so a master dark frame was 
constructed by combining darks recorded on all nights.

	Flat-field frames were obtained at the beginning of each night. 
Images of a dome spot were recorded with the flat-field lamps on and then off. The 
former frames contain information about pixel-to-pixel sensitivity variations and 
vignetting of the optical path that is required to 
flat-field the data, although thermal emission from uncooled objects 
along the optical path is also present. However, the latter frames 
contain information only about thermal emission from 
uncooled objects, and so a calibration image containing 
only flat-field information can be obtained by subtracting the 
images with lamps `off' from those with lamps `on'. The 
flat field frames constructed in this manner showed no detectable night-to-night 
variations, and so a single master flat was created for each filter using 
flat-field data obtained throughout the run. Photometric studies of globular 
clusters, which are simple stellar systems and so are ideal for 
measuring uncertainties in the flat-field, using CFHTIR data indicate that
the flat-field frames constructed in this manner track sensitivity variations across 
the field at roughly the 1\% level.

	A calibration frame that monitors signatures from thermal emission was 
constructed by median-combining flat-fielded images of `blank' fields 
that were observed periodically throughout the run. 
Flat-fielded images of each sky field in each filter 
were corrected for DC sky variations, normalized to a common integration 
time, and then median-combined to reject sky sources.

	The raw data were reduced using a standard pipeline for near-infrared 
imaging that corrects for additive (dark current, DC sky level, 
thermal emission, and interference fringes) and multiplicative 
(pixel-to-pixel sensitivity variations and vignetting of the optical path) effects 
using the calibration frames described above. The basic steps in the reduction process 
were: (1) dark subtraction, (2) the division by flat-field frames, (3) the 
subtraction of the DC sky level, which was measured on a frame-by-frame basis, 
and (4) the subtraction of the interference fringe and thermal emission calibration 
frame constructed for each filter. The individual processed images for each tile were 
spatially registered to correct for the dither offsets and then median-combined on a 
filter-by-filter basis. The combined images were trimmed to the region 
of common overlap, and the final $K'$ image of the southernmost tile is shown in 
Figure 2.

\subsection{GMOS Data}

	Broad-band $r'$ and $i'$ images of a field to the south of NGC 
55, the location of which is indicated in 
Figure 1, were recorded with the GMOS imaging spectrograph 
on GS during the nights of October 18 and 21, 2003. The detector mosaic in GMOS 
consists of three $2048 \times 4068$ EEV CCDs, with 
each pixel sampling 0.072 arcsec on a side. The 
detector was binned $2 \times 2$ during readout for these observations, and so the 
angular scale is 0.144 arcsec superpixel$^{-1}$. 
Crampton et al. (2000) describe the various GMOS components in detail.

	Six 150 sec exposures, recorded with a $2 \times 3$ 
point dither pattern to allow the gaps between the detector 
elements to be filled during processing, were obtained in each filter. 
A control field, located midway between NGC 55 and NGC 247 with $\alpha = 
00^{hr} 31^{min} 58.1^{sec}$ and $\delta = 
-29^{o} 54^{'} 57.8^{"}$ (E2000), was also observed with a total 
exposure time of 2700 sec per filter. This longer exposure time 
for the control field was needed to match observations of a field 
in the halo of NGC 247, which is the subject of a separate paper (Davidge 
2004, in preparation). The image quality in the NGC 55 and control field images is 
roughly 0.6 arcsec FWHM.

	Bias frames were recorded at various times throughout the two week 
observing run, and these were median-combined to create a master bias frame. 
Flat-field frames were constructed from unguided images of the twilight sky. Each 
twilight sky image was normalized to produce a mean flux level of unity, and the 
results were median-combined to suppress objects on the sky. The results 
track pixel-to-pixel sensitivity variations and vignetting across the science 
field at better than the 1\% level. Finally, an $i'$ fringe frame was constructed by 
median-combining individual images of the NGC 55 and control field frames. The 
fringes are suppressed at a level that is less than 1\% of their original amplitude.
Interference fringes are not evident in the $r'$ data, and so a fringe frame 
was not constructed for these data. 

	The data were reduced with a standard pipeline for CCD images that 
parallels in many respects the procedure described previously in \S 2.1 and 
utilizes the calibration frames discussed in the previous paragraph. The steps in the 
CCD reduction pipeline were: (1) the subtraction of the master bias frame, (2) the 
division by the flat-field frame, (3) the subtraction of the DC sky level from each 
exposure, and (4) for the $i'$ data, the subtraction of the fringe frame. 
The individual exposures for each field were spatially 
registered to correct for dither offsets and then
median combined. The combined images were trimmed to the area of common 
overlap, and the final $r'$ images are shown in Figures 3 (NGC 55) 
and 4 (Control field).

\subsection{Photometric Measurements and Analysis Strategy}

	The photometric measurements were made with the PSF-fitting program 
ALLSTAR (Stetson \& Harris 1988). Tasks in DAOPHOT (Stetson 1987) were used 
to identify stars, measure preliminary brightnesses, and construct PSFs. 
The photometric calibration of the CFHTIR data was set using observations 
of standard stars from Hawarden et al. (2001), while for the GMOS data this was done 
using standard stars from Landolt (1992). The brightnesses and colors of the Landolt 
standards were transformed into the SDSS photometric system using 
equations from Fukugita et al. (1996). Based on the scatter in the standard 
star measurements, the photometric zeropoints have an uncertainty 
of a few tenths of a magnitude in $J, H,$ and $K$ and less than 0.01 magnitudes 
in $r'$ and $i'$.

	Davidge (1998) discussed near-infrared photometry of a field in NGC 55 
that was observed with the CIRIM camera on the CTIO 1.5 meter telescope and 
that overlaps the southern end of the CFHTIR strip. 
Although there is a significant difference in angular resolution between 
the two datasets (FWHM = 1.5 -- 1.8 arcsec for the CTIO data, compared with 
1.0 arcsec for the CFHT data), the photometric measurements of the stars with 
$K < 16$ made with the two instruments are in reasonable 
agreement. In particular, the mean difference in $K$, 
in the sense CFHTIR -- CIRIM, is $\overline{\Delta K} 
= 0.18 \pm 0.13$, while for $J-K$ it is $\overline{\Delta (J-K)} 
= -0.05 \pm 0.06$. The quoted errors are the standard deviations about 
the mean. Crowding is more of an issue for stars in the CTIO images 
given the poorer angular resolution, and this will 
bias upwards brightnesses obtained from these data, which is consistent with 
the sign of $\overline{\Delta K}$.

	Completeness fractions and the random errors in the photometry were 
assessed by running artificial star experiments. Based on the CMDs constructed 
from the data and discussed in \S 3 and 4, artificial stars 
inserted into the CFHTIR data were assigned colors matching those of AGB stars, while 
artificial stars inserted into the GMOS data had colors appropriate for RGB stars. 
Artificial stars were considered to be recovered only if they were 
detected in two or more filters. These experiments indicate that 
the 50\% completeness level occurs at $K = 18$ in 
the densely populated regions near the NGC 55 disk plane, and $K = 19.5$ 
in the less crowded extraplanar regions. As for the GMOS data, 
the 50\% completeness level occurs near $i' = 25.5$ in the extraplanar regions.

	If two or more stars fall within the same angular resolution element then 
they will appear as a single object, and the artificial star experiments can 
be used to assess the incidence of such blends. Given the crowded nature of the 
NGC 55 disk almost all stars in the southernmost tile of the CFHTIR dataset 
are likely to be blends, but most of these involve 
a very faint star being combined with a bright object, so 
that the effect on the photometric properties of the brighter object are 
insignificant. However, if two sources having brightnesses within a few tenths of 
a magnitude of each other fall in the same resolution element then they will be 
detected as an object that is $\sim 0.5$ mag brighter than the more luminous 
of the two progenitors. The artificial star experiments predict that in the disk 
of NGC 55 about 3\% of detected objects with $K = 17$ are blends of objects 
having comparable brightness, while by $K = 18$ the blend fraction increases 
to 10\%. These fractions drop considerably in the extraplanar regions. 
As for the GMOS data, blending only becomes significant near the faint end, with 
10\% of the sources near $i' = 25$ being blends of objects with comparable brightness.

	The data discussed in this paper sample a range of environments. To 
facilitate the analysis and investigate the stellar contents of these environments, 
the galaxy was divided into the following regions: (1) 
the thin disk, which in turn was sub-divided into star clusters and 
the field, (2) the thick disk, and (3) a region that is
loosely referred to as `the halo'. The boundaries of the various regions 
are marked in Figure 1, and the rationale for placing the boundaries shown in Figure 1 
are discussed in the next paragraph. 

	There are two obvious star clusters in the CFHTIR data and the 
boundaries of these, indicated in Figure 2, were set by eye. Both of these 
clusters are sites of H$\alpha$ emission in the map constructed by T\"{u}llmann et 
al. (2003), and so are expected to be made up of young stars. The thin disk 
was defined to lie within two of the stellar disk vertical 
scale lengths measured by Kiszkurno-Koziej (1988). 
Because NGC 55 is inclined slightly, the foreground and background portions of 
the disk broaden the projected width of the galaxy. Assuming a disk radius of 
6 kpc, estimated from the POSS image, then a projected width of 2.2 arcmin on 
either side of the disk center line will include the thin disk regions of 
the near and far ends of the NGC 55 disk, and the thin disk 
boundaries shown in Figure 1 have this width. As for the thick 
disk, Dalcanton \& Bernstein (2002) found that this component tends to extend 
to at least 5 vertical disk scale lengths off of the disk plane in very 
late-type disk systems. After accounting for the inclination 
of NGC 55 in the same way as was done for the thin disk, the 
thick disk boundaries were defined to fall between 2.2 and 4.4 
arcmin off of the disk center line. Finally, the halo is defined simply as 
the region that falls beyond the outer boundary of the thick disk, although 
in \S 4 it is demonstrated that this region appears simply to be a low surface 
brightness extension of the thick disk. The stellar contents of the star 
clusters and thin disk field are discussed in \S 3, 
while the thick disk and halo regions are discussed in \S 4.

\section{RESULTS: THE YOUNG DISK OF NGC 55}

\subsection{The Thin Disk Field as Imaged with CFHTIR}

	The $(K, H-K)$ and $(K, J-K)$ CMDs of stars in the thin disk field 
are shown in the right hand panels of Figures 5 and 6. The $(K, J-K)$ CMD in Figure 6 
is in many respects similar to the CMDs of the NGC 55 fields studied by Davidge (1998), 
one of which overlaps with the area observed with the CFHTIR. In 
particular, the prominent vertical sequence in the Davidge (1998) CMDs have 
a peak $K-$band brightness and $J-K$ color that matches the CMD in the right hand 
panel of Figure 6. However, the CFHTIR CMD goes deeper and has less scatter 
than the Davidge (1998) CMDs, due to the larger 
aperture of the CFHT and better image quality.

	The $(K, J-K)$ and $(K, H-K)$ CMDs both broaden towards the faint end. The 
standard deviation along the color axis for stars with $K$ between 17.75 and 18.25 is 
$\pm 0.26$ mag in $H-K$ and $\pm 0.36$ mag in $J-K$. For comparison, the artificial 
star experiments predict that random photometric errors introduce a scatter of 
$\pm 0.09$ mag in $H-K$ and $\pm 0.15$ mag in $J-K$ at $K = 18$. Given that the 
scatter in both CMDs exceeds the random photometric error predictions, it 
thus appears that the broadening of the CMDs near the faint end is due to a real spread 
in stellar content. Later in this section it is demonstrated that this spread 
is due to AGB stars having a range of ages.

	The objects along the vertical sequence in the right hand panels of Figures 5 
and 6 with $K < 16.5$ are RSGs, while the sources on this sequence with $K > 16.5$ 
are evolving on the AGB (see below). The nature of the stars in the thin disk field 
can be investigated further by examining their broad-band near-infrared 
spectral-energy distributions (SEDs). This is done in the upper panel of Figure 7, 
which shows the $(J-H, H-K)$ two-color diagram (TCD) of stars in the thin 
disk with $K < 17.5$. The majority of stars have SEDs that are suggestive of 
M giants, although the SEDs of some are consistent with K and M dwarfs, 
indicating that they may be foreground objects. 
The stars with $H-K > 0.4$ tend to scatter around the 
LMC LPV sequence, and there are some objects with $J-K > 1.6$, 
which is a color range dominated by C stars in the LMC (Hughes \& Wood 1990).

	The narrow-band CO and K continuum observations were used 
to measure the CO index of the brightest stars in the southernmost tile of the 
CFHTIR strip, and the $(K, CO)$ CMD and $(J-K, CO)$ TCD of stars in the 
thin disk field are shown in Figures 8 and 9. The stars fall along a single 
sequence in the CMD, and tend to have CO indices that are consistent with them being 
M giants, although some have CO indices that are indicative of RSGs.
The brightest star in the thin disk field has a CO index near 0, suggesting 
that it may be a foreground dwarf.

	The $(M_K, J-K)$ CMD of the thin disk is shown in the right hand panel 
of Figure 10. A distance modulus of $\mu_0 = 26.5$, based 
on the RGB-tip brightness measured from the GMOS data (\S 4.2), has been 
adopted. Foreground reddening is negligible towards NGC 55 (Burstein \& Heiles 
1984), and so no reddening correction was applied. Also shown in Figure 10 are 
isochrones from Girardi et al. (2002). Given the overall similarities 
between NGC 55 and the LMC, a metallicity of Z = 0.008 ([Fe/H] $\sim -0.3$), which is 
comparable to the peak in the metallicity distribution function of giants in the 
LMC (Larsen, Clausen, \& Storm 2000), might be appropriate for NGC 55. 
In fact, HII regions in NGC 55 have oxygen abundances that are similar to those 
in the LMC, although NGC 55 may be deficient in nitrogen with respect to the 
LMC (Webster \& Smith 1983). T\"{u}llmann et al. (2003) argues that the typical disk 
metallicity of NGC 55 is Z = $0.007^{+0.002}_{-0.001}$, based on the 
central HII region. Thus, the isochrones plotted in Figure 10 have Z = 0.008.

	The log(t) = 8.15 isochrone is significant as it marks the onset of 
the AGB, and this isochrone comes close to matching the color 
of the AGB plume in NGC 55. The predicted peak AGB brightness also falls 
close to a discontinuity in the number counts along the vertical sequence in 
the CMD, although matching the AGB-tip brightness is problematic, as 
many bright AGB stars are photometric variables, with the brightest being observed near 
the peak of their light curves. The peak AGB brightness also has a slight 
metallicity sensitivity, such that M$_K^{AGBT} = -9.8$ if Z = 0.008, which is
0.1 mag brighter than the predicted peak Z=0.004 brightness. In any event, it appears 
that a significant number of AGB stars in the thin disk of NGC 55 have progenitor 
masses that are close to that needed for the AGB to form. These are 
accompanied by a sizeable population of fainter AGB stars with colors and 
brightnesses that are consistent with log(t) $\geq 9$, indicating that 
this portion of NGC 55 has been an area of active star formation for at least 
the last 1 Gyr.

	The comparison with the isochrones in Figure 10 indicate that stars with M$_K 
< -10$ are RSGs, and it is evident that the RSG sequence in the thin disk extends to 
M$_K = -11.5$. This peak RSG brightness is roughly consistent with what is 
seen in other star-forming disk-dominated systems, although there is considerable 
galaxy-to-galaxy scatter. Late-type galaxies with 
integrated brightnesses comparable to NGC 55 have peak M$_K$ stellar brightnesses 
between --11 and --12 (e.g. Rozanski \& Rowan-Robinson 1994), while the 
brightest RSGs in the Magellanic Clouds have M$_K < -11$ and $J-K = 1$ (Elias, 
Frogel, \& Humphreys 1985). 

	The age of the brightest RSGs can be estimated from 
their luminosities. The $K-$band bolometric correction for 
RSGs with $J-K = 1$ is BC$_K = 2.75$ (Elias et al. 1985), 
and so RSGs with M$_K = -11.5$ have M$_{bol} = -8.7$, or log(L/L$_{\odot}) 
= 5.4$. This corresponds to stars with progenitor masses near 25 
M$_{\odot}$, or ages of $\sim 8$ Myr based on the models computed by 
Fagotto et al. (1994). The bright RSGs are thus the result of very recent star 
formation throughout the disk of NGC 55. 

	The integrated near-infrared broad-band colors 
of the thin disk were measured to permit comparisons with other galaxies. 
There are prominent dust lanes in the area separating Clusters 
1 and 2, and so the integrated colors were measured in a 
semi-circular annulus with a width of 10 arcsec that samples the disk to 
the north, south, and west of Cluster 1, but avoids the region 
between the two clusters. Caution must be 
exercised in making integrated photometric measurements of this nature, as  
the presence of a handful of very bright stars may unduly influence 
aperture photometry. Therefore, rather than compute the mean flux in the 
area described above, the median flux was computed instead. 

	The location of the thin disk field on the TCD is shown in 
Figure 11, along with models for simple stellar systems from Leitherer et al. 
(1999). The location of the thin disk field on the TCD 
is appropriate for an intermediate-age system. This is not an unexpected result
given the preponderance of AGB stars in the near-infrared CMDs of the 
thin disk field. The integrated colors of SB(s)m systems in the 2MASS Large Galaxy 
Atlas (Jarrett et al. 2003), as computed from the total brightness measurements, 
are also shown in Figure 11. The SB(s)m galaxies 
form a sequence that cuts diagnonally across Figure 11, and the 
NGC 55 thin disk field falls along this sequence. This comparison indicates that the 
SED of the thin disk field, and thus the stellar content, is consistent with 
what is seen in other galaxies of similar morphological type. 

\subsection{The Thin Disk as Imaged with GMOS}

	The outer regions of the thin disk skirt the north east corner 
of the GMOS field, and the $(i', r'-i')$ CMD of this portion of the GMOS dataset 
is shown in the left hand panel of Figure 12. The 
$(i', r'-i')$ CMD of the corresponding region of the control 
field is shown in left hand panel of Figure 13. The 
control field contains a mixture of foreground stars and background galaxies, 
with the latter dominating near the bright end.

	The GMOS data sample intrinsically fainter stars than the 
CFHTIR data because they were recorded with a much larger telescope and 
have better image quality. The $(i', r'-i')$ CMD of the thin disk is 
dominated by RGB and AGB stars, with the RGB-tip near $i' = 23.1$ (\S
4.2); the properties of the RGB in the thin disk are not discussed 
here because of the crowded nature of this part of the GMOS image, and a 
detailed discussion of the RGB in NGC 55 is deferred to \S 4.2. However, 
it is worth noting that the color of the RGB in the thin disk portion of 
the GMOS data is similar to what is seen in the thick disk, 
suggesting similar metallicities. 

	The $(i', r'-i')$ CMD of the thin disk portion of the GMOS field 
is populated by stars spanning a range of ages, 
indicating that this portion of the thin disk has formed stars over an 
extended period of time. A population of blue stars 
with $r'-i' = -0.3$ and $i' > 21.5$ is present, and these likely belong to 
the youngest populations in this portion of NGC 55. Applying the 
transformation equations from Fukugita et al. (1996) indicates that 
the brightest blue stars have $V \sim 21$, or M$_V \sim -5.5$. These stars 
are thus near the bright end of the main sequence (e.g. Humphreys \& McElroy 1984), 
and so may be very massive and have ages consistent with, or even younger 
than, the RSGs discussed in \S 3.1 and 3.2.

	There are two AGB sequences in the GMOS thin disk CMD. One 
sequence forms a plume with $r'-i' \sim 0.4$ that extends roughly from the 
RGB-tip to $i' = 21$, while the other has $i' > 22$, and bends towards larger 
$r'-i'$ colours, becoming difficult to track when $r'-i' > 1.5$. 
The nature of these AGB sequences is investigated in Figure 14, which shows
the $(M_{i'}, r'-i')$ CMD of the thin disk and the corresponding 
portion of the control field. A distance modulus of $\mu_0 = 26.5$ has been 
assumed (\S 4.2). Also shown are Z=0.001 and Z=0.008 isochrones from Girardi et 
al. (2002), which were transformed into the SDSS filter system using relations 
from Fukugita et al. (1996).

	The log(t$_{yr}$) = 8.15 isochrone roughly follows the 
vertical AGB plume. Given the number of objects in 
the control field CMD, it is likely that the 
bulk of stars with M$_{i'} \sim -6$ are AGB stars in NGC 55, 
and not background or foreground contaminants. As for the fainter AGB 
sequence, while the log(t$_{yr}$) = 10 models with Z=0.001 and Z=0.008 have 
very different colors, it is clear that the upper 
envelope of the fainter AGB stars on the $(M_{i'}, r'-i')$ CMDs 
is consistent with log(t$_{yr}$) = 10.0. Therefore, the thin disk of NGC 55 contains 
stars that formed early in the life of the galaxy. 
Stars fall between the Z=0.008 log(t$_{yr}$) = 8.15 and 10.0 sequences, indicating that 
star formation continued during intermediate epochs. 

	The bolometric LF of AGB stars provides another means of probing the 
star-forming history of the thin disk of NGC 55. To construct such a LF from the 
GMOS data, AGB stars were selected based on (1) brightness, such that $i' 
< 23.1$, so as to be brighter than the RGB-tip, and (2) color, such that $r'-i' 
> 0.2$ to avoid main sequence stars and include the blue edge of the AGB plume. 
While the thin disk of NGC 55 contains RSGs, these will 
have $i'< 20$, and no such objects are seen in the GMOS field.

	Following Davidge (2003), the $r'-i'$ colors and $i'$ brightnesses were 
transformed into $R-I$ and $I$, where $R$ and $I$ 
are in the Kron-Cousins system, using relations from 
Fukugita et al. (1996) and the relation between $V-I$ and $R-I$ for 
solar neighborhood giants from Bessell (1979). Bolometric magnitudes were 
then computed using $I-$band bolometric corrections for AGB stars 
from Bessell \& Wood (1984). This procedure was also applied to objects in the thin 
disk portion of the control field, and the result was subtracted from the LF 
constructed from the NGC 55 thin disk field. The final bolometric AGB LF 
of the thin disk, corrected for objects in the control field, is shown in Figure 15. 

	The LF does not follow a single power-law, and there is a clear break near 
M$_{bol} = -4.6$. The majority of stars with M$_{bol} > -4.6$ are on the AGB 
immediately above and to the right of the RGB-tip 
in the $(i', r'-i')$ CMD. The location of these stars on the CMD is consistent with 
log(t$_{yr}) = 10$, indicating that they have `old' ages (see above). 
The break in the LF thus suggests that the star formation rate in this region 
of the NGC 55 disk dropped at some point during the past $\sim 10$ Gyr, and 
that a large fraction of the stars in the thin disk formed during early epochs.

	Davidge (1998) investigated the bolometric LF of two areas in the disk of 
NGC 55, and it is of interest to see if a break occurs 
in those LFs near M$_{bol} = -4.6$. Unfortunately, after adjusting the LFs in Figure 11 
of Davidge (1998) to the distance modulus adopted here, incompleteness becomes 
significant when M$_{bol} > -5$, and so these data do not go faint enough to determine 
if a break is present. Pritchet et al. (1987) also investigated the 
bolometric LF of AGB stars in NGC 55. After adjusting their 
LF to the distance modulus adopted here, the faint limit of the Pritchet et 
al. (1987) LF is M$_{bol} \sim -5$, and so these data also do not go 
faint enough to sample the discontinuity seen in the LF 
constructed from the GMOS data.

\subsection{Clusters 1 and 2}

	The $(K, H-K)$ and $(K, J-K)$ CMDs of Clusters 1 and 2 are shown in 
the left hand and middle panels of Figures 5 and 6. The cluster CMDs 
show more scatter and are shallower than the thin disk CMDs, due 
to the higher stellar densities in the clusters. Indeed, Cluster 1 is more 
crowded than Cluster 2, and the CMDs of the former show a greater amount 
of scatter than the latter. It is likely that some of the 
stars in the cluster CMDs may be interlopers from the thin disk field. This 
being said, the brightest stars in the clusters 
are $\sim 0.5$ mag in $K$ fainter than in the thin disk field, and this 
leads to only a modest difference in age (see below).

	The $(J-H, H-K)$ TCD of stars with $K < 17.5$ in both clusters 
is shown in the upper panel of Figure 7. As was the case in the thin disk, 
the majority of stars within the cluster boundaries have near-infrared 
SEDs that are similar to those of solar neighborhood M giants. There is 
considerable scatter among stars in both clusters with $H-K > 0.3$, which is the 
color interval occupied by LPVs in the LMC. Some stars in Cluster 1 fall in a 
region of the $(J-H, H-K)$ TCD where $J-K > 1.6$, indicating that they 
may be C stars.

	The $(K, CO)$ CMDs of stars in Clusters 1 and 2 are 
shown in the left hand and middle panels of Figure 8, 
while the $(J-K, CO)$ TCD of stars with $K < 16$ in Clusters 1 and 2 is 
shown in the upper panel of Figure 9. The brightest stars in Clusters 1 and 2 
define a relatively tight sequence in Figure 8, with $\overline{CO} = 
0.22$ when $K < 16$. The stars in Cluster 2 
fall in a region of the $(J-K, CO)$ TCD that is occupied by M giants, while many of the 
stars in Cluster 1 have CO indices that are consistent with them being 
RSGs. There is one star in Cluster 1 with $J-K \geq 1.5$ that has a CO 
index that is comparable to those of Milky-Way C stars. 
Finally, the brightest star in Cluster 2 has CO = 0.0, which is 
much weaker than the CO indices of the majority of stars in either cluster. 
Based on this weak CO index, it is likely that this object is a foreground 
star, and not a real member of the cluster.

	The $(M_K, J-K)$ CMDs of Clusters 1 and 2 are shown in the left hand and middle 
panels of Figure 10. The ideal means of measuring the age of Clusters 1 and 2 
would be to use the brightness of the main sequence turn-off. While some bright 
blue stars are seen in the CMDs of both clusters 
near M$_K = -9.5$, there are not enough objects on which to base a 
reliable age determination, and these could be 
blue supergiants. Therefore, as with the thin disk field, the 
ages of Clusters 1 and 2 are estimated from the properties of the RSGs.
The RSGs with M$_K = -11$ have M$_{bol} = -8.2$, or log(L/L$_{\odot}) 
= 5.2$. This corresponds to stars with progenitor masses near 20 M$_{\odot}$, 
or an age of $\sim 10$ Myr based on the Fagotto et al. (1994) models. 
Hence, although the brightest RSGs in the thin disk field are $\sim 0.5$ mag 
brighter in $K$ than their counterparts in Clusters 1 and 2, this leads to 
only a 2 Myr difference in age. The young age inferred 
for these clusters is consistent with them being sites of H$\alpha$ 
emission (T\"{u}llmann et al. 2003).

	The integrated near-infrared photometric properties of Clusters 1 and 2 
can be used to check the age estimated from the RSGs. 
Following the procedure described in \S 3.1 for the thin disk field, the integrated 
colors of the clusters were computed from median flux measurements 
to suppress the effects of individual bright stars.
The background was measured in a 2 arcsec wide 
annulus immediately around each cluster.

	The location of Clusters 1 and 2 on the $(J-H, H-K)$ TCD is shown in Figure 11. 
A comparison with models from Leitherer et al. (1999) indicates that both clusters have 
near-infrared SEDs that are consistent with them being simple stellar 
systems with log(t$_{yr}) \sim 6.8$ and A$_V$ between 1 and 2 magnitudes; 
thus, the integrated near-infrared colors of Cluster 1 and 2 are broadly consistent 
with the age inferred from the RSGs. 

	Davidge (2004) detected a population of young clusters near the center of 
the star-forming galaxy NGC 3077. The young NGC 3077 clusters discovered by Davidge 
(2004), which are also shown in Figure 11, form a relatively 
tight sequence on the TCD that roughly parallels the reddening vector. 
Clusters 1 and 2 have near-infrared SEDs that are similar to those of 
the NGC 3077 clusters, but with 1 -- 2 magnitudes less extinction in $V$. 

\subsection{EHR 1}

	T\"{u}llmann et al. (2003) discovered two star-forming regions that have 
projected distances of 1 -- 1.5 kpc from the NGC 55 disk plane. One of 
these, EHR 1, falls within the boundaries of the thin disk region and was imaged with 
CFHTIR. Individual sources were not resolved in this object. However,
the integrated brightness of EHR 1 in a 2.5 arcsec radius aperture is 
$K = 16.1$, with $(H-K) = 0.72$ and $(J-H) = 0.92$. The near-infrared 
SED of EHR 1 is consistent with it being a very young system, 
and this is demonstrated in Figure 11, which shows the location of EHR 1 
on the near-infrared TCD. EHR 1 falls on the NGC 3077 young cluster
sequence in Figure 11, and a comparison with the Leitherer et al. (1999) models 
indicates that EHR 1 has an age of a few Myr, with A$_V = 5 - 6$ magnitudes.

\section{RESULTS: THE THICK DISK AND HALO}

\subsection{The CFHTIR Strip}

	The $(K, H-K)$, and $(K, J-K)$ CMDs of the thick disk and halo 
portions of the CFHTIR strip are shown in Figures 16 and 17. There is a 
concentration of stars in the thick disk CMDs when $K > 18.5$, which is due to 
stars evolving on the AGB. In constrast, the CMDs of the halo region lack an 
obvious concentration of AGB stars.

	The numbers of stars with $K < 18$ in the thick disk and halo CMDs are similar, 
suggesting that the objects at this brightness likely do not belong 
to NGC 55. This is consistent with the near-infrared SEDs of these objects.
The $(J-H, H-K)$ TCD of stars with $K < 17.5$ in the thick 
disk and halo is shown in the lower panel of Figure 7. While 
the majority of stars in the thin disk have near-infrared SEDs that 
are consistent with them being evolved giants, and hence are 
likely at the distance of NGC 55, many of the bright sources in the 
thick disk and halo regions fall systematically below the giant sequences on 
the TCD. In fact, the stars with $H-K < 0.4$ occupy a portion of the 
TCD that suggests they are K and M main sequence stars, and so 
are likely foreground objects. Some of the brighter 
objects in the halo field have obvious extended morphologies, 
indicating that they are background galaxies.

	The $(M_K, J-K)$ CMDs of the thick disk and halo are shown in 
Figure 18. Based on the number of objects in each CMD, 
an excess population of stars in the thick 
disk with respect to the halo region only occurs when 
M$_K > -7.5$, and so M$_K = -7.5$ is adopted as the AGB-tip brightness in the thick 
disk. The majority of resolved RGB stars in this portion of NGC 55 have metallicities 
between Z = 0.001 and 0.004 (\S 4.2), and so the isochrones in Figure 18 
have these metallicities. Given an AGB-tip brightness of M$_K = -7.5$, the 
isochrones in Figure 18 indicate that the resolved stars in the thick disk 
have an age that is close to log(t$_{yr}$) = 10. The stellar contents of the thin 
disk and the thick disk thus appear to differ, in that the latter 
lacks the young and intermediate age populations seen in the former. 

\subsection{The GMOS Data}

	The $(i', r'-i')$ CMDs of the thick disk and halo portions of the NGC 
55 GMOS field are shown in the middle and right hand panels of Figure 12; 
the CMDs of the corresponding regions of the control field are shown in Figure 
13. The number of objects with $i' < 22$ in the control field 
is comparable to that in the thick disk and halo CMDs, 
indicating that many of the sources with $i' < 22$ in the thick disk and halo 
are probably foreground and background objects. 
The bright main sequence stars and vertical AGB plume that were 
conspicuous in the thin disk are thus abscent in detectable numbers, and 
this is consistent with what is seen in the CFHTIR data (\S 4.1). 

	Stars evolving on the RGB and AGB dominate 
the faint end of the thick disk and halo CMDs. The giant 
branch in the thick disk is much wider than predicted solely from photometric errors. 
Indeed, the standard deviation in $r'-i'$ for stars with $i'$ between 22.75 and 23.25 
is $\pm 0.12$ mag, whereas the artificial star experiments predict a scatter due to 
photometric errors of only $\pm 0.03$ magnitudes. Thus, there is a real variation in 
stellar content among bright giants, and later in this section this is shown to be due 
to a spread in metallicity. 

	The brightness of the RGB-tip is a standard candle, and 
the RGB-tip brightness was measured from the thick disk data. The 
number counts in the thick disk CMD drop noticeably near $i' = 23$, indicating 
that the RGB-tip occurs close to this brightness. A better estimate of the 
RGB-tip brightness can be obtained from the $i'$ LF of RGB stars in the thick disk, 
which is shown in Figure 19. The final LF was constructed by restricting the 
sample to objects with $r'-i'$ between --0.5 and 1.5 and then 
correcting for objects falling in the same color interval in the corresponding region 
of the control field. 

	The conventional means of measuring the brightness of the RGB-tip is to 
convolve the LF with a Stobel edge-detection kernel (e.g. Lee, Freedman, 
\& Madore 1993). However, given the significant spray of AGB stars 
above the RGB-tip, it was decided to apply 
a different technique, which judges the position of the RGB-tip based on the 
departure from the power-law nature of the LF, as defined by RGB stars below 
the RGB-tip.

	The NGC 55 LF shown in the lower panel of Figure 19 follows a power-law at the 
faint end. A power-law was fit to the LF bins between $i' = 23.5$ and 25.0, which is 
an interval where the uncertainties in individual bins are modest and 
the artificial star experiments predict that 
incompleteness is not significant. A least squares fit to the thick 
disk LF entries in this brightness interval gives an exponent 
$x = 0.22 \pm 0.01$, which is within the 
range seen in globular clusters (e.g. Davidge 2000). The 
fitted power-law is compared with the data in the lower 
panel of Figure 19. The LF departs significantly from the fitted relation 
at $i' \sim 23.1$, and this is adopted as the brightness of the RGB-tip.

	Lee et al. (1993) calibrated the RGB-tip as a standard candle 
in the I filter. Following Davidge et al. (2002), the 
relations computed by Fukugita et al. (1996) were used to transform the $i'$ 
brightness of the RGB-tip into I, with the result that I$^{RGBT} 
= 22.5$ for NGC 55. Adopting M$_{I}^{RGBT} = -4$ (Lee et al. 1993), which 
holds for populations with [Fe/H] $< -0.7$, then the distance to NGC 55 
computed from the RGB-tip is $\mu = 26.5$. No correction is applied for 
foreground extinction, as this appears to be small (Burstein \& Heiles 1984). 

	With the distance to NGC 55 established from the brightness of the 
RGB-tip then the metallicity of the thick disk and halo can be estimated from 
the colors of RGB stars in the $(i', r'-i')$ CMD. In Figure 20 the $(M_{i'}, 
r'-i')$ CMD of RGB stars in the thick disk and halo are compared with 
log(t$_{yr}$) = 10 sequences from Girardi et al. (2002), transformed 
into the SDSS system using relations from Fukugita et al. (1996).
The isochrones match the RGB-tip brightness and shape of the upper RGB 
in both the thick disk and halo regions. The majority of RGB stars in the 
thick disk appear to have a metallicity between Z = 0.001 and Z = 0.003, although 
there are also stars that may be as metal-poor as Z = 0.0001. 
The RGB stars in the halo have a similar range of metallicities to those in the thick 
disk, and so there is no evidence for a 
metallicity gradient off of the disk plane. 

	There is a spray of objects above the RGB-tip in the thick disk 
CMD, and these are analogous to the AGB stars 
in the $(K, J-K)$ CMD that were discussed in \S 4.1. The AGB stars in the 
left hand panel of Figure 20 are nicely bracketed by the log(t$_{yr}$) 
= 10 isochrones with Z=0.001 and Z = 0.004; in fact, the upper envelope of the 
AGB population in the left hand panel of Figure 20 is well matched 
by a line that connects the ends of the Z=0.001 and Z=0.004 
log(t$_{yr}$) = 10.0 AGB sequences. The inferred old age for the AGB stars 
is consistent with what was established from the CFHTIR data in \S 4.1. 

	What is the uncertainty in the age estimate determined from the 
comparisons in Figure 20? The age estimate 
is subject to uncertainties (1) in the models, (2) in the ability to locate 
the upper envelope of the AGB population on the CMD due to photometric 
variability, and (3) in the assumed distance to NGC 55. The uncertainties in 
the models are difficult to assess, and an error of $\pm 0.1$ mag is 
(arbitrarily) assigned to this component. The scatter in the upper envelope of the 
thick disk AGB sequence on the CMD suggests that stellar variability may contribute 
a $\pm 0.1$ mag uncertainty. The errors in the distance estimates 
are more easily quantifiable, and are likely on the order of $\pm 0.1$ mag, 
which is one half of the binning interval in Figure 19. If all three components are 
combined in quadrature then the total uncertainty in M$_{i'}^{AGBT}$ is $\pm 
0.2$ mag. The Z = 0.001 AGB isochrones are more-or-less 
vertical in the $(M_{i'}, r'-i')$ CMD, and hence are well suited 
for assessing how an error in the AGB-tip brightness propogates into an uncertainty 
in the age estimate. The Z=0.001 sequences from Girardi et al. (2002) indicate that a 
0.2 mag change in M$_{i'}^{AGBT}$ translates into a 0.1 dex change in 
log(t$_{yr}$) near log(t$_{yr}$) = 10, and so a $\pm 0.1$ uncertainty is assigned to 
log(t$_{yr}$) estimated from the AGB-tip brightness. A lower limit to the age of the 
dominant stellar component in the envelope surrounding NGC 55 is thus 8 Gyr, while an 
upper limit is 13 Gyr.

	While an AGB sequence is not seen above the RGB in the halo 
CMD, this does not mean that these stars are abscent; rather, they may have a 
density that is sufficiently low to make their detection difficult. In fact, 
the $i'$ LF of stars in the halo, corrected for sources in the control field, 
contains a modest population of stars that extend up to 0.6 mag in $i'$ above 
the RGB-tip, indicating that AGB stars brighter than the RGB-tip do occur in 
this field. Based on the similar metallicities and age properties inferred from 
the CMDs, it appears that the halo and thick disk regions as defined 
here are not physically distinct entities; rather, they are part of the same 
structure.

	Dalcanton \& Bernstein (2002) find that the envelopes that surround 
their sample of undisturbed, bulgeless, edge-on disk galaxies 
have many properties in common. In particular, there is a distinct break between 
the main disk bodies and the surrounding envelopes, while the integrated 
colors of the thick disks indicate that the envelopes 
tend to have a metallicity that is comparable to the thin disk, and
contain stars that formed $\sim 6$ Gyr in the past. Finally, the envelopes 
also tend to be homogeneous stellar systems, in the sense that there is 
no evidence for age or metallicity gradients at moderately large distances off 
of the disk plane (e.g. Figure 10 of Dalcanton \& Bernstein 2002). 

	While the properties of the envelope around NGC 55 are 
reminiscent of those in the Dalcanton \& Bernstein (2002) sample, 
few envelopes in that sample have [Fe/H] as low as $-1$, and ages 
as old as 10 Gyr. The Dalcanton \& Bernstein (2002) analysis 
relies on integrated photometric measurements, and so may be 
affected by the well known age-metallicity degeneracy. 
This degeneracy works in the sense that higher metallicities may be 
misinterpreted as older ages, with the result that if metallicity is 
overestimated then ages will be underestimated. Thus, if the metallicities 
of the envelopes in the Dalcanton \& Bernstein (2002) sample were fixed at 
lower values then older ages would be predicted. This would then give results 
that are consistent with the NGC 55 envelope properties established here from 
resolved stars.

\section{DISCUSSION \& SUMMARY}

	Moderately deep red and near-infrared images that were obtained with 
the CFHTIR imager on the CFHT and the GMOS imaging spectrograph on GS have been 
used to probe the stellar content of the nearby galaxy NGC 55. The data are 
used to investigate the stellar content along the disk plane and in the 
extraplanar region using evolved red stars as probes. The results of 
this investigation provide insight into the past history of the galaxy.

\subsection{The Recent Star-Forming History of NGC 55}

	In \S 3.1 it was shown that the integrated near-infrared SED of the disk near 
the center of NGC 55 is similar to the SEDs of other Magellanic irregular galaxies, and 
is consistent with significant recent star formation. In fact, the thin disk field 
contains a population of RSGs that have brightnesses consistent with very young ages, 
indicating that the region near the center of NGC 55 has been the site of 
large scale star formation within the past 8 Myr. There are also two large 
clusters that have ages near 10 Myr. The CFHTIR data samples 
the disk near the main light concentration in NGC 55. The presence of very 
young stars in this portion of the NGC 55 disk may not be surprising if the 
main light concentration is a bar viewed end-on (de Vaucouleurs 1961), as the ends of 
bars in Magellanic-type galaxies tend to be sites of very recent star formation 
(Elmegreen \& Elmegreen 1980).

	A substrate of luminous AGB stars is seen throughout the thin disk, 
and comparisons with models from Girardi et al. (2002) indicate that these objects 
span a range of ages. A prominent feature in the CMDs 
constructed from both the CFHTIR and GMOS data is a plume that appears to contain 
stars with masses close to that needed to commence 
evolution on the AGB. A plume of stars with a similar color and peak brightness is seen 
in the CMDs of the two NGC 55 fields studied by Davidge (1998). One of the 
Davidge (1998) fields samples a region of the NGC 55 disk that is, like the regions 
observed with GMOS, well offset from the optical center of the galaxy. Thus, the 
stars on the AGB plume occur throughout the disk of NGC 55.

	The detection of large numbers of stars with masses near the RSG -- AGB 
evolutionary transition provides clues into the star-forming history of NGC 55 during 
recent epochs. One possibility is that NGC 55 is being viewed at a fortuitous 
time following an isolated episode of star formation when the RSG -- AGB 
transition just happens to be occuring. However, it is more likely that the 
detection of significant numbers of stars with masses close to those needed 
to evolve on the AGB indicates that there has been 
vigorous star formation during the past 0.1 -- 0.2 Gyr. 
In fact, data from other sources indicate that the SFR in NGC 55 during the 
past 0.1 - 1 Gyr has been at least as high as that at the present day, and 
that it is more likely that the star formation rate (SFR) of NGC 55 has 
recently dropped. A number of indicators support a present-day SFR of at most a few 
tenths of a M$_{\odot}$ year$^{-1}$. The FIR luminosity and radio continuum emission 
from NGC 55 suggest that the present-day SFR is 0.3 M$_{\odot}$ year$^{-1}$ 
(Dettmar \& Heithausen 1989), while the integrated H$\alpha$ flux of NGC 55 is 
consistent with a SFR of 0.16 M$_{\odot}$ year$^{-1}$ (Ferguson et al. 1996). The 
SFR computed from x-ray flux measurements ranges between 0.06 and 
0.18 M$_{\odot}$ year$^{-1}$, based on equations 14 and 15 of Ranali, Comastri, 
\& Setti (2003). Recent FIR flux measurements from the Spitzer Space Telescope yield a 
SFR of 0.22 M$_{\odot}$ year$^{-1}$ (Engelbracht et 
al. 2004). These various estimates of the present-day SFR are 
significantly lower than what is computed from the blue luminosity, which is a 
measure of the SFR averaged over the past Gyr, and which predicts a SFR of 1.6 
M$_{\odot}$ year$^{-1}$ in NGC 55 (Dettmar \& Heithausen 1989). Further evidence that 
the SFR has dropped during the past $60 \times 10^7$ years comes from the high SNe rate 
needed to explain the extended x-ray emission around the galaxy (Oshima et al. 2003).
There may have been brief lulls in star-forming activity, as T\"{u}llmann et al. (2003) 
argue that breaks in the SFR may be needed to allow gas in the extraplanar region to 
cool sufficiently for stars to form, as has happened in EHR 1 and 2.

	The stellar content throughout the thin disk of NGC 55 
appears to be well mixed. For example, the distribution of 
the $J-K$ colors of stars with $K$ between 16.0 and 17.5 does not change with 
distance from the disk plane; a gradient in the color distribution would 
be expected if mean age changed with height above the disk plane. 
The ratio of objects with SEDs consistent with M 
giants to stars with SEDs similar to LMC LPVs also does not change 
with distance above the disk plane in the thin disk region. 
Adopting $H-K = 0.3$ as the dividing line between M giants 
and LPVs in the TCD, then the ratio of M giants to LPVs is 11/8 $= 1.4 \pm 0.6$ 
in the portion of the thin disk between 1.1 and 2.2 arcmin from the 
disk plane, compared with 162/125 $= 1.3 \pm 0.2$ for stars within 1.1 arcmin 
of the disk plane.

	The similarity in the stellar contents of the regions probed with the 
CFHTIR, which samples the disk plane, and GMOS, which samples the edge of 
the thin disk, further suggests that the disk of 
NGC 55 has experienced a remarkably uniform star-forming history. However, 
it should be kept in mind that NGC 55 is viewed nearly edge-on, and 
so any given site line samples stars from a large 
range in galactocentric distances. A high inclination can thus introduce an 
apparent uniformity in stellar content, as the star-forming history 
is smoothed over large portions of the galaxy. This also makes it easier to detect 
rare features in CMDs, such as the onset of the AGB, as the projected density of 
objects is higher than in other orientations. 

	We close this portion of the discussion by noting that the 
contribution made by C stars to the total AGB luminosity 
peaks near log(t$_{yr}) \sim 8.7$ in moderately metal-poor systems
(Maraston 1998); therefore, a large C star population 
might be expected in NGC 55 based on the star-forming history inferred from various 
sources, including the CFHT and GMOS data. However, Figure 9 of Pritchet et al. 
(1987) suggests that the C star frequency in NGC 55 may be relatively low 
given the integrated brightness of the galaxy. It is thus worth noting 
that candidate C stars with $K < 17.5$ are seen in the $(J-H, H-K)$ TCD of thin 
disk stars. These C stars may well be the tip of the iceberg, and their 
presence suggests that a rich C star population remains to be 
discovered in the disk of NGC 55.

\subsection{The Extraplanar Region and the Early Evolution of NGC 55}

	The CFHTIR and GMOS data sample the extraplanar stellar content 
of NGC 55, and both datasets support an old age for this region of the galaxy; 
more specifically, based on the peak brightnesses of stars on the AGB there is no 
evidence for a young or intermediate age component 
outside of the region that is defined here as the thin disk. 
This result is perhaps surprising given the presence of gas and dust 
outside of the disk plane (e.g. Ferguson et al. 1996). Of course, the extraplanar 
region is not completely devoid of young stars, as there are some pockets of 
recent star formation (T\"{u}llmann et al. 2003). Nevertheless, it appears 
that extraplanar star-forming regions like EHR 1 and EHR 2 have not contributed 
significantly to the stellar content of the envelope surrounding NGC 55.

	The metallicities of EHR 1 and 2 suggest that they formed 
from material that did not originate in the thin disk (T\"{u}llmann 
et al. 2003). In fact, the metallicities of these HII regions 
are in excellent agreement with the mean metallicity 
among RGB stars in the extraplanar region. This suggests 
that the age-metallicity relation in the extraplanar region may have been flat 
over a large fraction of the age of the Universe, after an initial period of 
rapid enrichment (see below). This could occur if the envelope 
has not been the site of large scale star formation for the past 10 Gyr or if it is 
unable to retain processed material. Another possibility is that there has been 
infall of metal-poor material at a rate that has fortuitously maintained the mean 
metallicity of the gas in the extraplanar region at a constant value.

	The results of this study have implications for understanding the early 
evolution of NGC 55 in the context of the hierarchal formation model, although whether 
the envelope formed from the disruption of an early proto-disk or from the violent 
relaxation of proto-galactic gas clouds remains a matter of speculation. 
The absence of obvious metallicity gradients indicates that the envelope formed 
in a highly coherent manner or has been well mixed since its formation.
It is perhaps worth noting that the mean metallicity of the envelope around NGC 
55 coincides with the point where the dynamical properties of the globular cluster 
system in the Milky-Way changes from orbits that are indicative of a 
pressure-supported system to one that has a significant rotational component 
(Zinn 1985), and this provides indirect evidence for a disk origin. 
Efforts to study the shape of the envelope will help constrain the origin of 
this structure. If the envelope originated from the disruption of a protodisk 
then it should not have a markedly flattened distribution. 

	The issue of how the envelope formed notwithstanding, that the RGB stars in the 
extraplanar region are old and typically have [Fe/H] $\sim -1$ indicates that they 
likely formed from material that experienced rapid early enrichment. This is broadly 
consistent with the age-metallicity relation of the LMC during early epochs, 
which de Vaucouleurs (1961) argues is a galaxy that is structurally similar 
to NGC 55. The LMC age-metallicity relation climbs from [Fe/H] $< -2$ to --1 over 
a 3 Gyr period during early epochs (Hill et al. 2000), and an age dispersion 
of this size would affect the AGB-tip brightness in $K$ in NGC 55 by only $\sim 
0.2$ mag (\S 4.2), making it difficult to detect. If the age-metallicity relations 
of NGC 55 and the LMC are similar then the absence of stars with [Fe/H] $> -0.8$ would 
indicate that there should be no stars with ages less than $\sim 10$ Gyr in the 
envelope around NGC 55, and this is consistent with the age inferred from the AGB-tip. 

	The chemical content of stars in the envelope around NGC 55
will provide additional clues into the origin of this structure. While there are as yet 
no abundance measurements of stars in the outer envelope of NGC 55, measurements 
of this nature have been made for metal-poor giants in the LMC, and 
the results are of interest in the context of the early evolution of Magellanic 
irregular galaxies. Hill et al. (2000) find that the metal-poor giants in the LMC 
fall along the trend between [O/H] and [Fe/H] defined by Galactic giants, 
suggesting that the metal-poor stars in the LMC may have formed in an environment 
similar to that from which the Galactic halo formed. However, the number of 
LMC giants in the Hill et al. (2000) sample is tiny, amounting to only four objects. 

	The globular cluster content of NGC 55 may present a problem for 
models in which the envelope around NGC 55 formed as the result of an early, violent 
collapse. If globular clusters form in supergiant molecular clouds (e.g. Harris \& 
Pudritz 1994) that are associated with the major collapse events that are 
expected to produce rapid chemical enrichment, then a number of globular 
clusters might be expected in NGC 55 given the age and metallicity properties 
of stars in the envelope. Indeed, the LMC contains ten globular 
clusters with ages older than 10 Gyr (e.g. Geisler et al. 1997), and the majority 
of these are metal-poor, with [Fe/H] $< -1$ (e.g. Hill et al. 2000). 
It appears that NGC 55 does not contain a large cluster population, and may have 
a globular cluster content that differs from that of the LMC. Da 
Costa \& Graham (1982) found three cluster candidates, one of which was identified as 
a young globular cluster, while Liller \& Alcaino (1983) found five candidate 
clusters. However, the nature of the clusters detected in these early surveys is 
in doubt, as Beasley \& Sharples (2000) obtained spectra of a 
number of globular cluster candidates and concluded that 
the vast majority of these are background galaxies, 
with only one being a bona fide globular cluster.

	The origin of the envelope around NGC 55 notwithstanding, the 
absence of a large population of young or intermediate age stars in the envelope 
indicates that the disk of NGC 55 has not been disrupted by 
major merger activity for a large fraction of the Hubble 
time; the conditions for a stable disk to form in this galaxy were 
evidently in place early on. In fact, NGC 55 is in a 
relatively isolated environment. It is one of four galaxies that are 
on the near side of the Sculptor group, and the largest member of this 
mini-group is NGC 300. The crossing time of this four galaxy ensemble is almost 19 
Gyr (Karachentsev et al. 2003); consequently, the number of galaxy-to-galaxy 
interactions is expected to be small. Nevertheless, there are hints that 
NGC 55 may have been involved in recent interactions. 
The HI disk of NGC 55 is warped (Hummel et al. 1986), while NGC 55 
is a barred galaxy, and bars are difficult to form without an external trigger, 
such as tidal interactions (e.g. Bottema 2003). This being said, the 
bar of the LMC contains a large population of stars with 
ages of at least a few Gyr (Ardeberg et al. 1997; Smecker-Hane et al. 
2002), suggesting that the bars in Magellanic irregular galaxies can be long-lived, 
and so may not be indicative of recent interactions.

\subsection{The Distance to NGC 55}

	The RGB-tip occurs near $i' = 23.1$ in the thick disk of NGC 55, 
indicating that the distance modulus of this galaxy is 26.5. This distance 
measurement is likely not affected by dust internal to NGC 55, as the 
brightness of the RGB-tip does not change with distance off of the disk, and any dust 
will likely be concentrated close to the disk plane. The 
distance modulus of NGC 55 measured from the RGB-tip is larger than that 
computed previously from other indicators (\S 1). Nevertheless, given the 
distances to other nearby galaxies in this part of the sky (Karachentsev 
et al. 2003), the distance estimated from the RGB-tip still makes 
NGC 55 one of the nearest members of the Sculptor group.

\acknowledgements{Sincere thanks are extended to the anonynmous referee, whose comments 
greatly improved the manuscript.}

\clearpage

\begin{figure}
\figurenum{1}
\epsscale{1.0}
\plotone{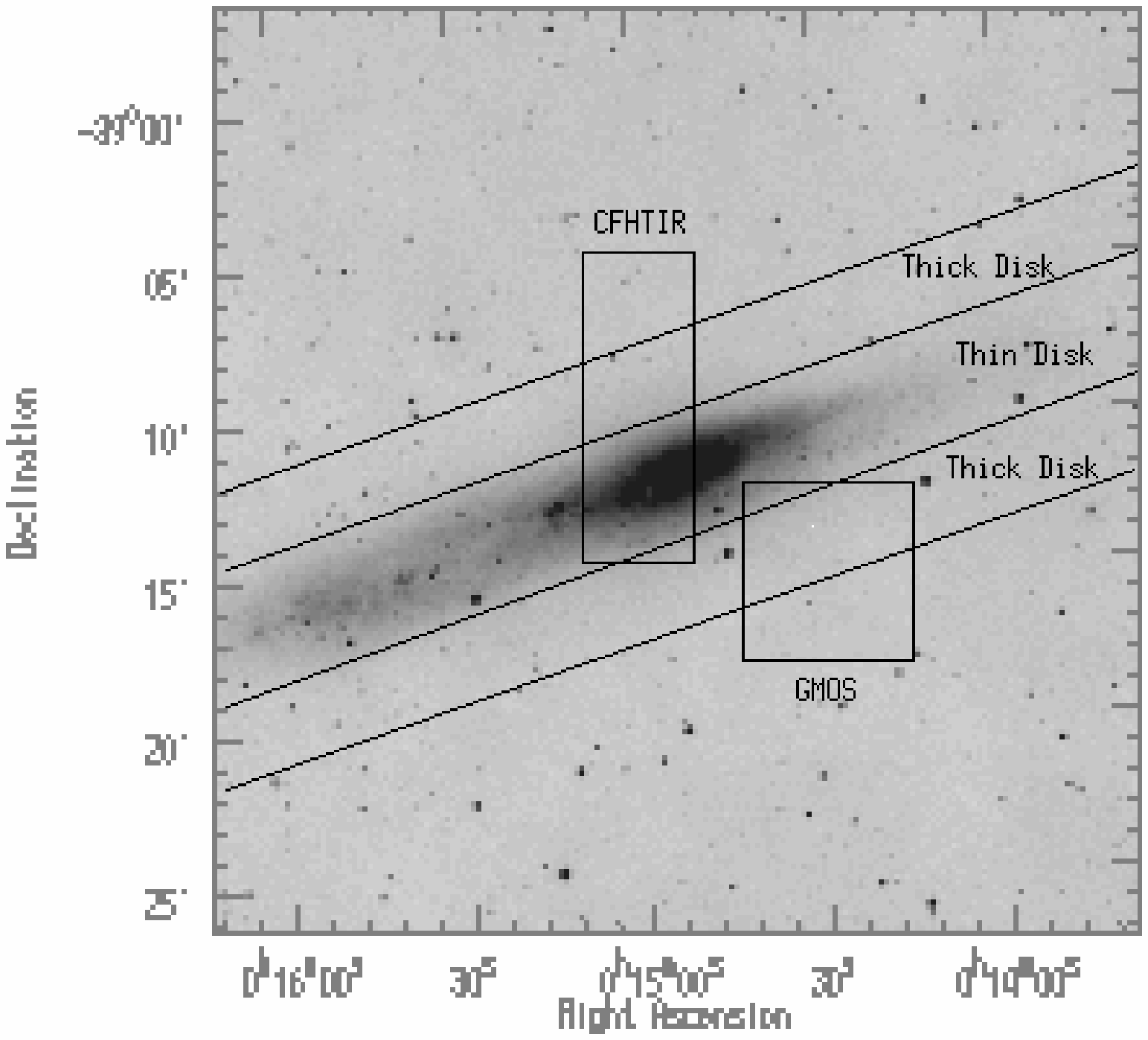}
\caption
{A $30 \times 30$ arcmin section of the Digitized Sky Survey 
showing the areas imaged with the CFHTIR and GMOS. East is to the left, 
and North is at the top. The lines mark the boundaries of the thin disk, 
thick disk, and halo regions as defined in \S 2.3.}
\end{figure}

\clearpage

\begin{figure}
\figurenum{2}
\epsscale{1.0}
\plotone{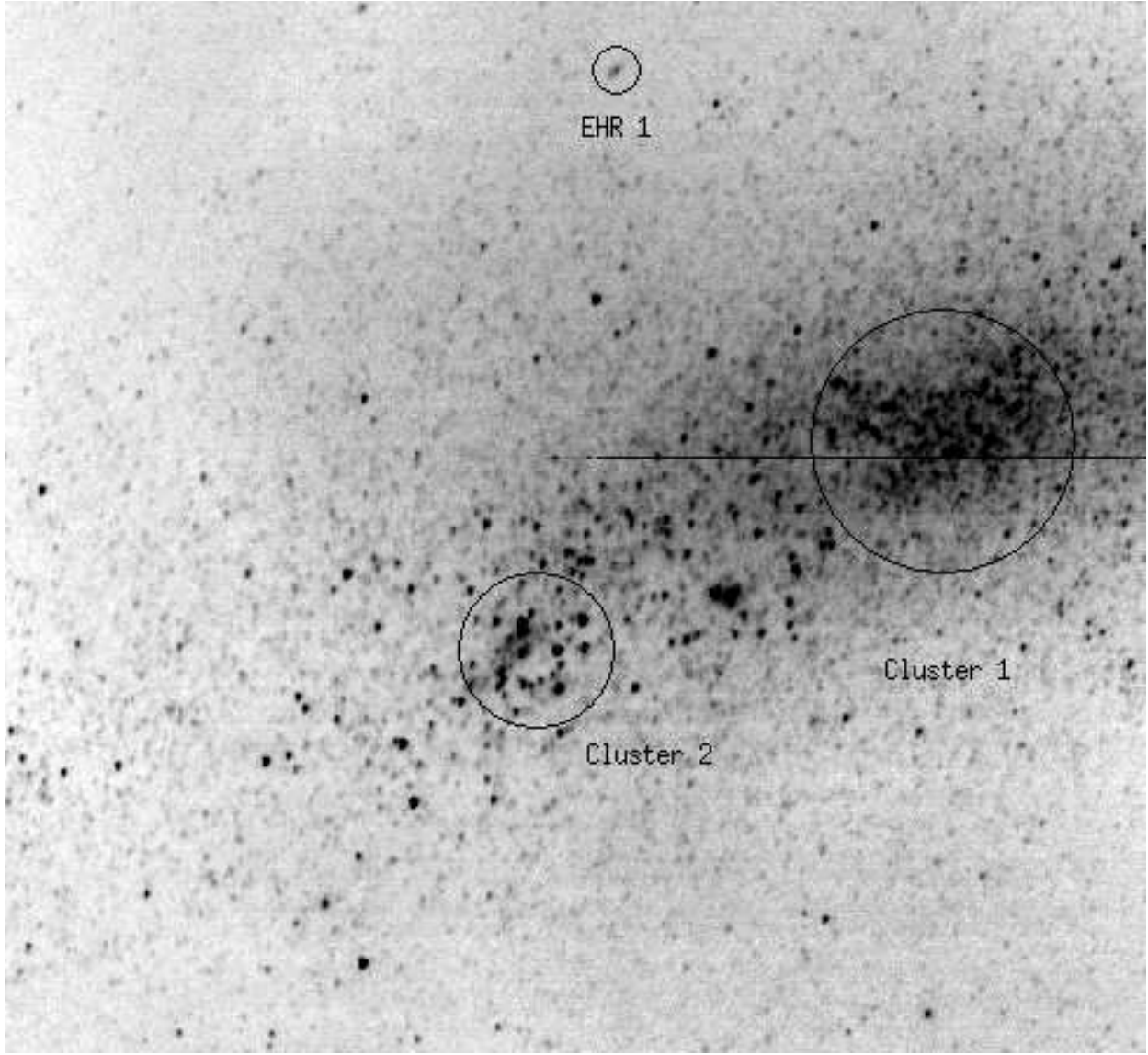}
\caption
{The final $K'$ image of the southernmost tile in the CFHTIR 
mosaic. A $3.6 \times 3.3$ arcmin field is shown. The two 
clusters that are discussed in \S 3 are indicated, as is the extraplanar 
HII region EHR 1. The horizontal line crossing through Cluster 1 is an 
artifact of the boundary between detector quadrants.
East is to the left, and North is at the top.} 
\end{figure}

\clearpage

\begin{figure}
\figurenum{3}
\epsscale{1.0}
\plotone{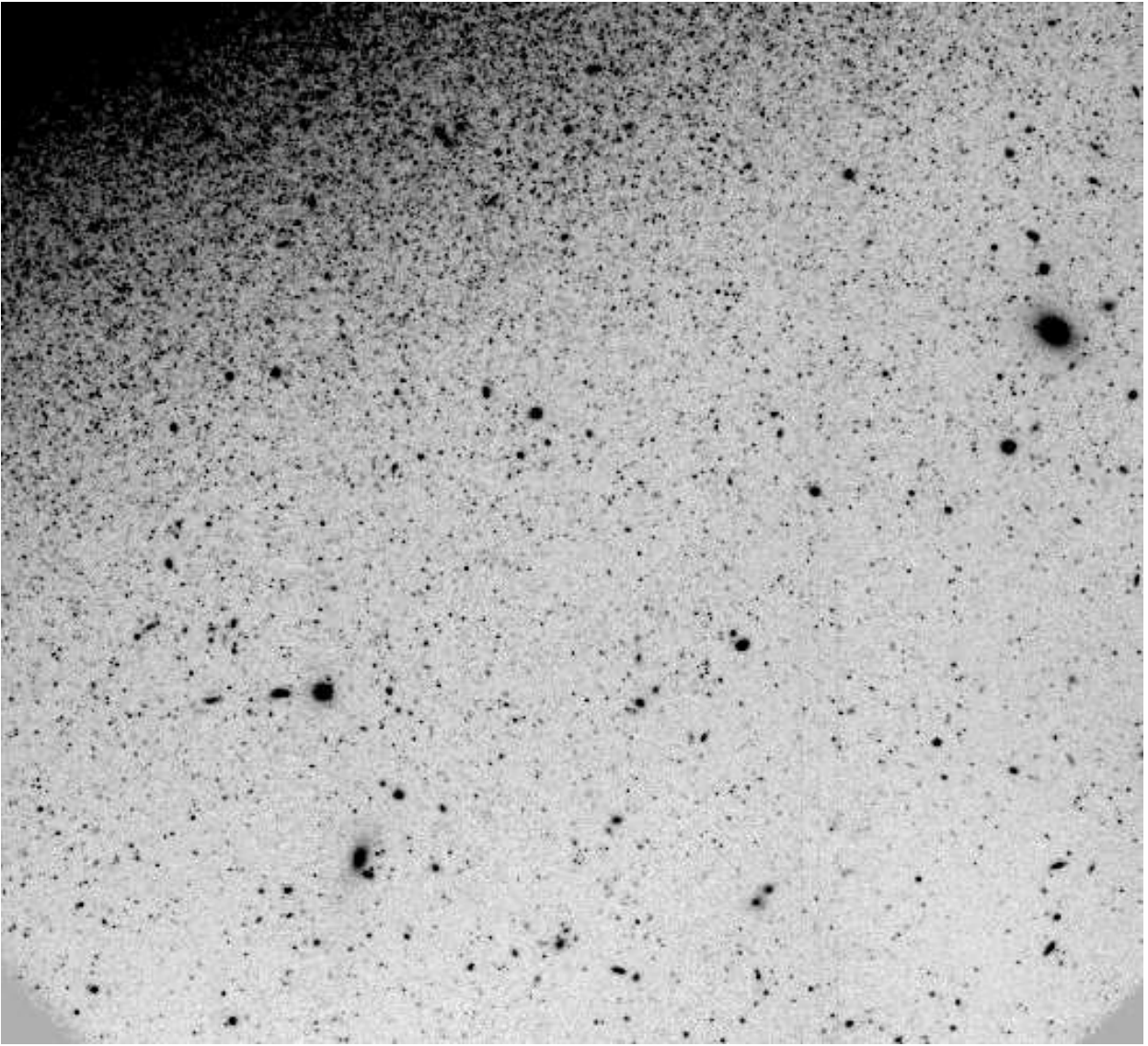}
\caption
{The final $r'$ image of the NGC 55 GMOS field, which covers 
$4.8 \times 4.8$ arcmin. The outer regions of the thin disk can be seen 
in the upper left hand corner. North is at the top, and East is to the left.}
\end{figure}

\clearpage

\begin{figure}
\figurenum{4}
\epsscale{1.0}
\plotone{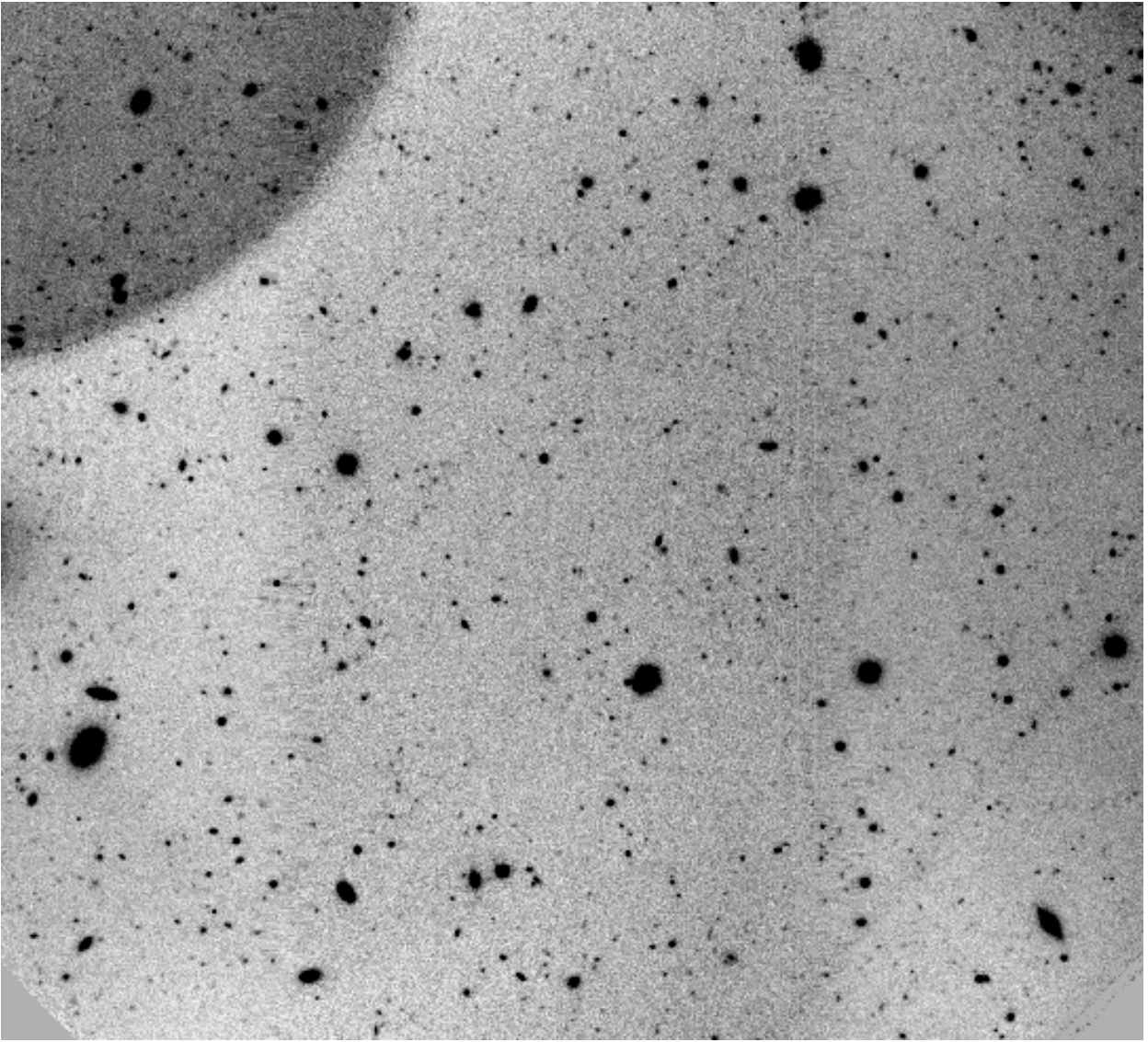}
\caption
{The final $r'$ image of the GMOS control field, which covers $4.8 \times 
4.8$ arcmin. The slight brightening of the background sky level in 
the upper left hand corner is an artifact of a bright star that is 
off of the field. North is at the top, and East is to the left.}
\end{figure}

\clearpage

\begin{figure}
\figurenum{5}
\epsscale{1.0}
\plotone{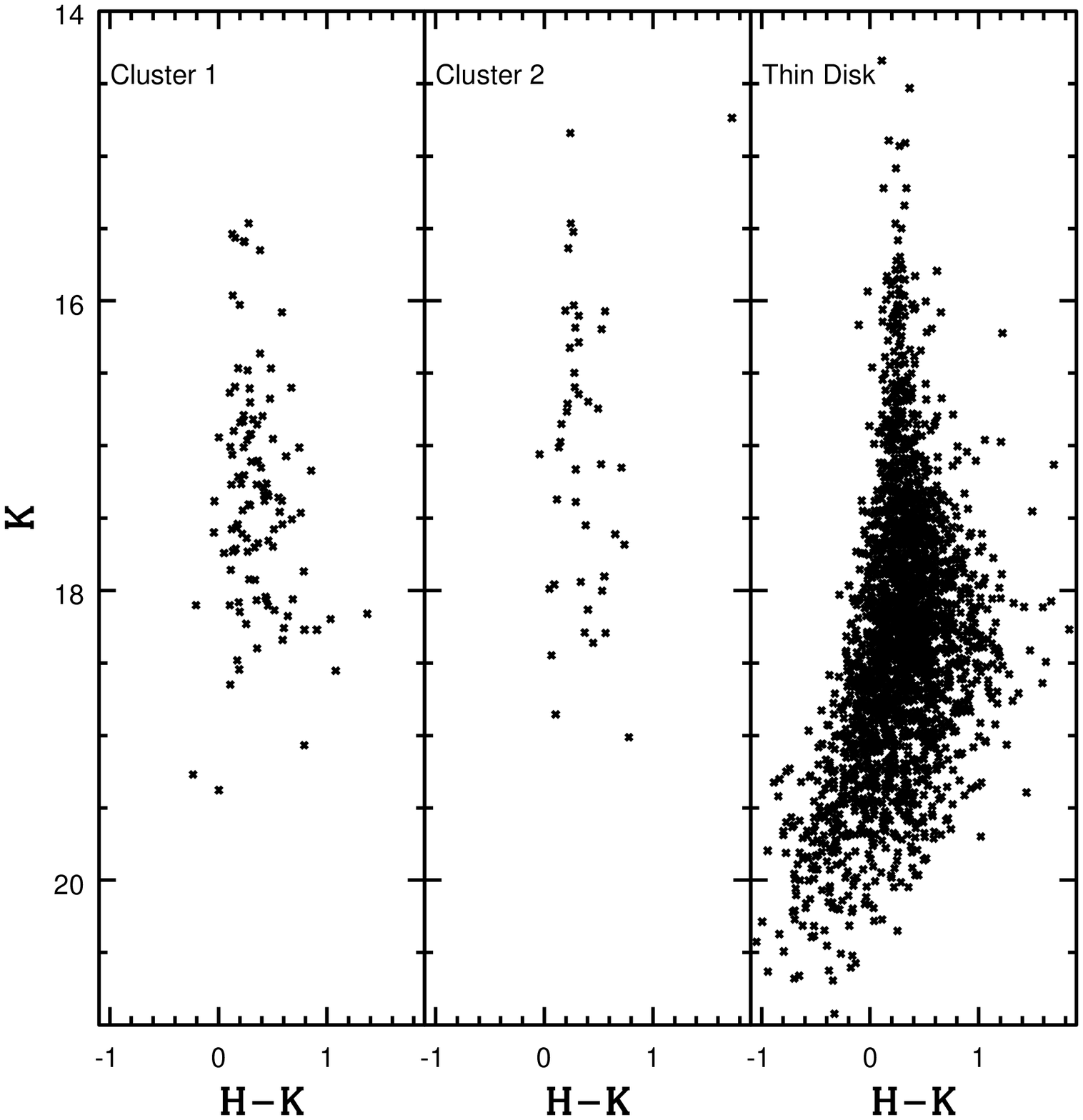}
\caption
{The $(K, H-K)$ CMDs of Cluster 1, Cluster 2, and the thin disk field. 
The vertical sequence in each CMD is dominated by AGB stars 
when $K > 16.5$ and RSGs when $K < 16.5$.}
\end{figure}

\clearpage

\begin{figure}
\figurenum{6}
\epsscale{1.0}
\plotone{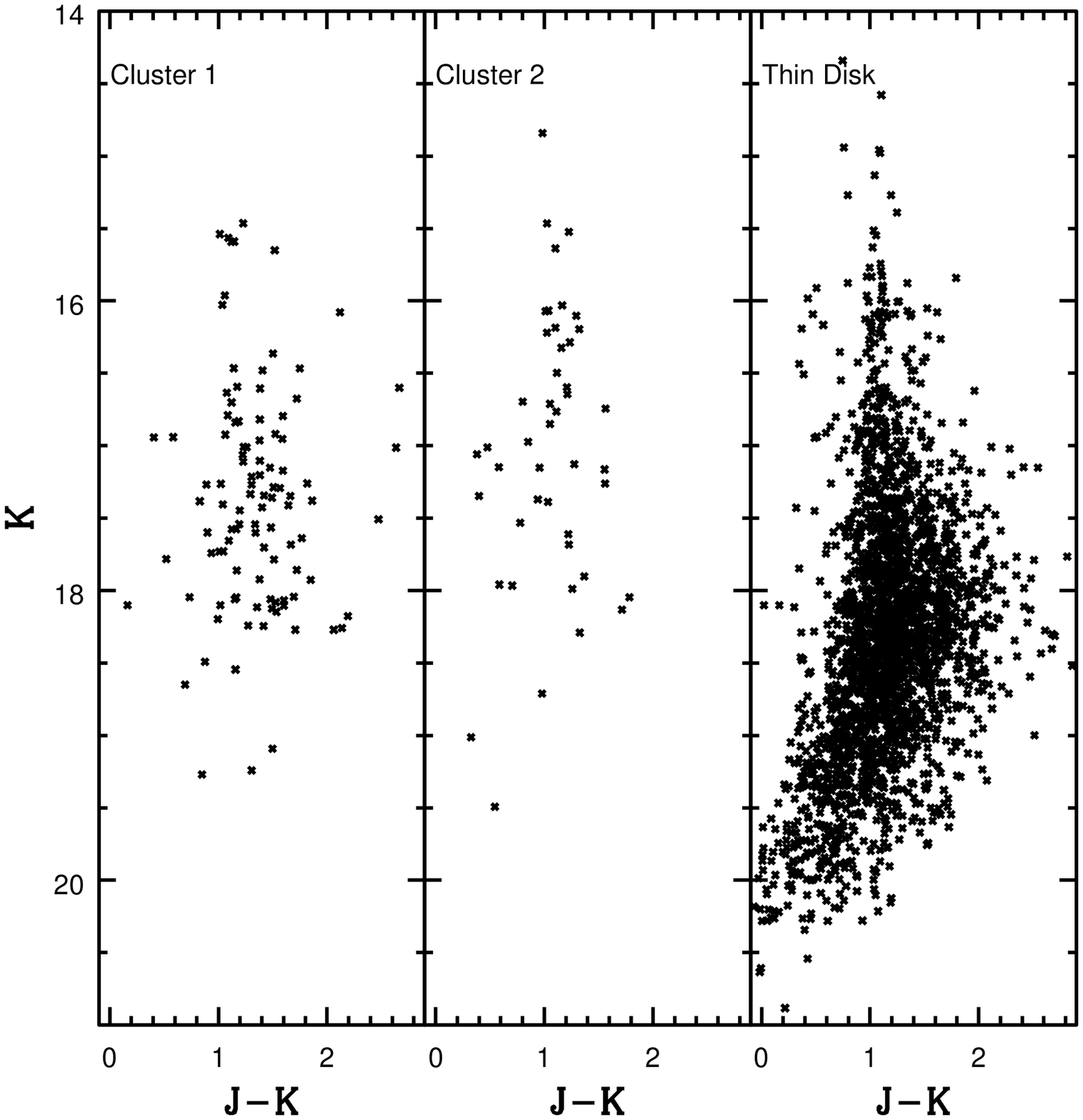}
\caption
{The $(K, J-K)$ CMDs of Cluster 1, Cluster 2, and the thin disk field. 
The vertical sequence in each CMD is dominated by AGB stars when 
$K > 16.5$, and RSGs when $K < 16.5$.}
\end{figure}

\clearpage

\begin{figure}
\figurenum{7}
\epsscale{0.9}
\plotone{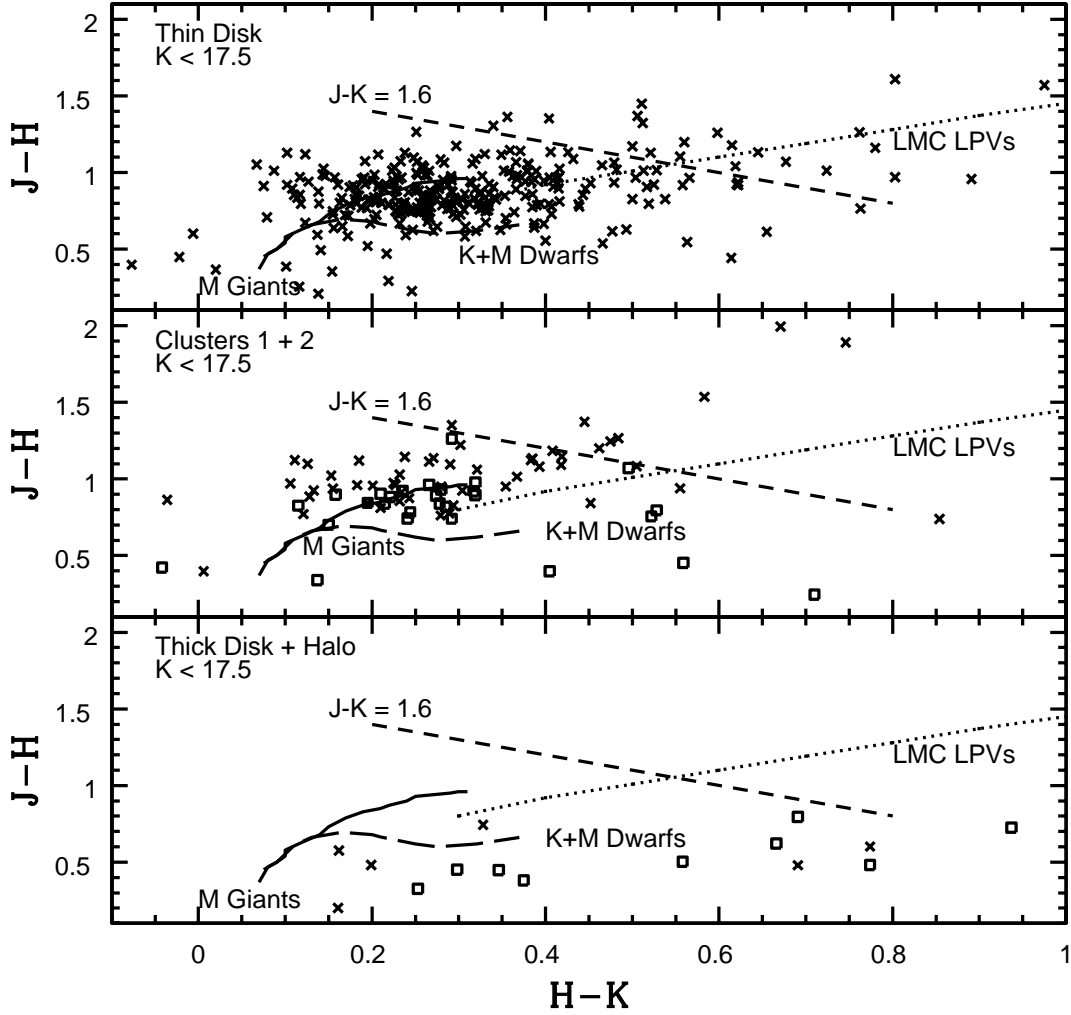}
\caption
{The $(H-K, J-H)$ TCD of stars with $K < 17.5$ in the thin disk field 
(top panel), Clusters 1 and 2 (middle panel), and the thick disk and halo 
(lower panel). Stars in the thin disk field, Cluster 1, and the thick disk are 
plotted as crosses, while stars in Cluster 2 and the halo are shown as open 
squares. The solid line is the sequence for solar neighborhood M 
giants from Bessell \& Brett (1988), while the dotted line is the 
locus of LMC LPVs from Hughes \& Wood (1990) 
and the long dashed line is the locus of K and 
M dwarfs from Bessell \& Brett (1988). The short dashed line indicates $J-K = 
1.6$; Hughes \& Wood (1990) find that LPVs in the 
LMC with $J-K < 1.6$ tend to be M giants, while those with $J-K > 1.6$ 
tend to be C stars.} 
\end{figure}

\clearpage

\begin{figure}
\figurenum{8}
\epsscale{1.0}
\plotone{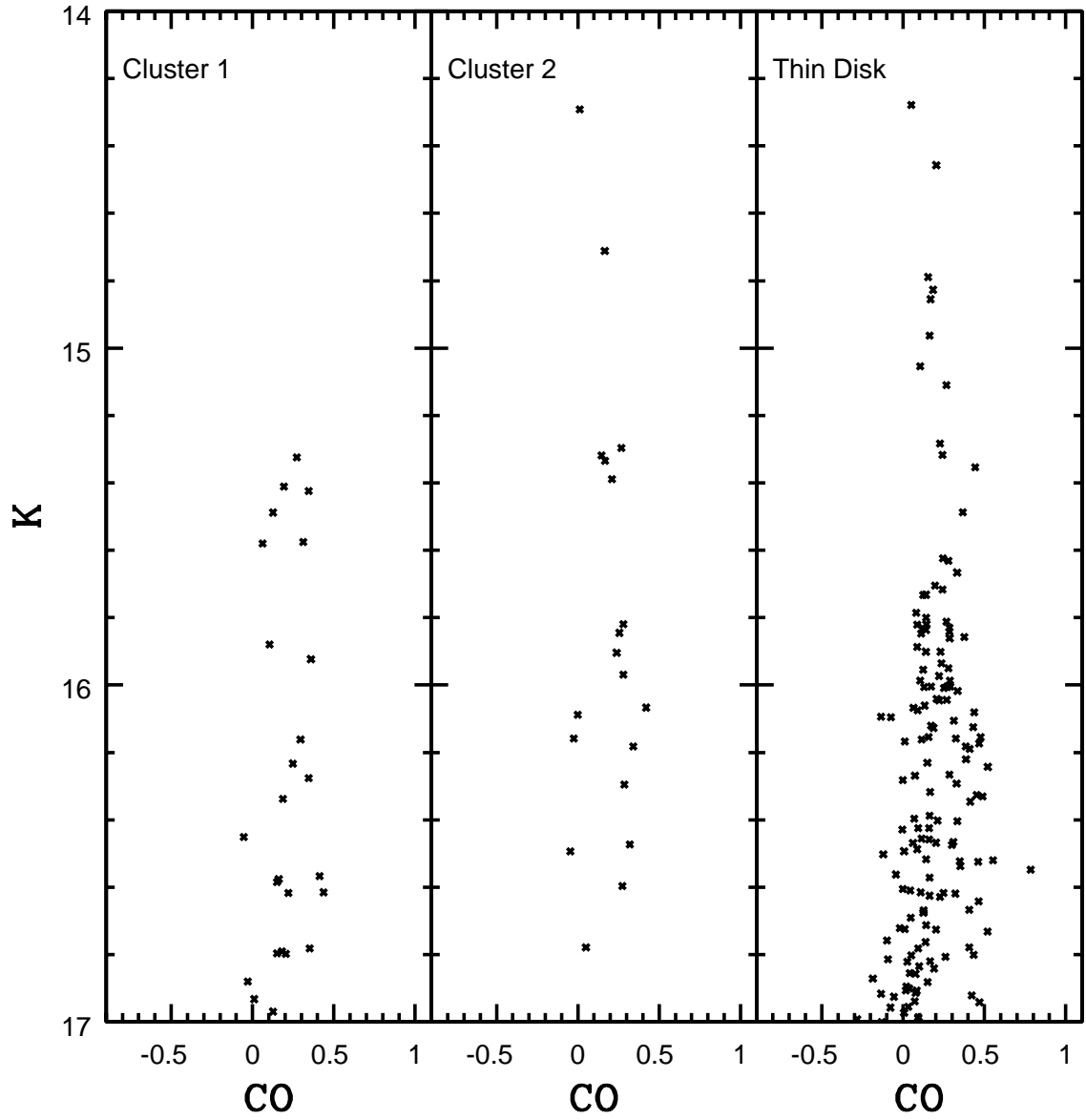}
\caption
{The $(K, CO)$ CMDs of Cluster 1, Cluster 2, and the thin disk field. 
The vertical sequences on the CMDs are 
dominated by RSGs and AGB stars. Note that the brightest star in Cluster 
2 and the thin disk field has a very weak CO index when compared with 
other stars, suggesting that it is a foreground object.}
\end{figure}

\clearpage

\begin{figure}
\figurenum{9}
\epsscale{1.0}
\plotone{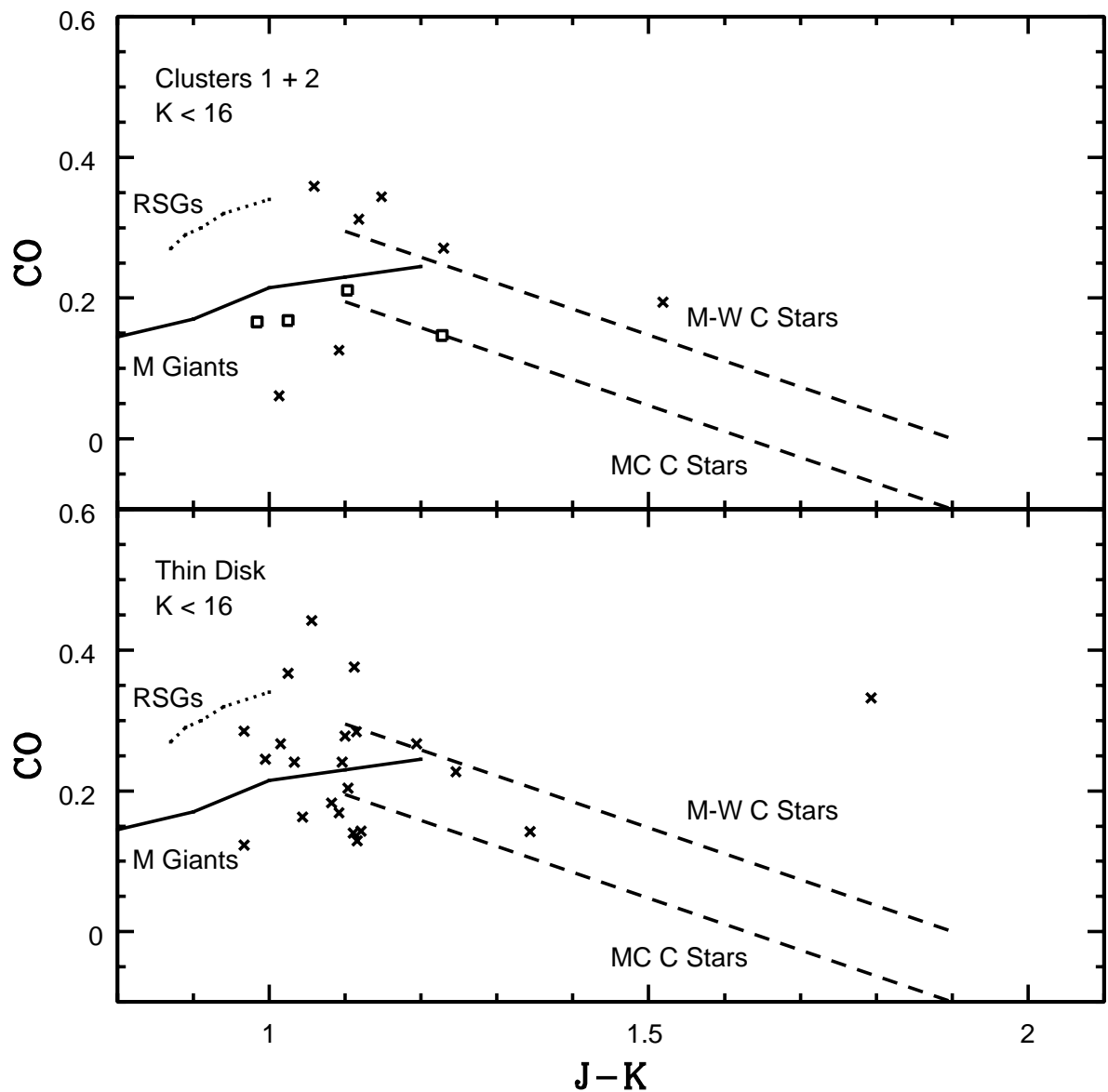}
\caption
{The $(CO, J-K)$ TCD for stars with $K < 16$ in Clusters 
1 and 2 and the thin disk field. Stars in Cluster 1 and the thin disk 
field are plotted as crosses, while stars in Cluster 2 are shown as open 
squares. The solid line shows the locus of Galactic M giants from Frogel et 
al. (1978), while the dotted line shows the sequence for LMC RSGs from 
Elias, Frogel, \& Humphries (1985). The dashed lines show the C 
star sequences for the Magellanic Clouds and the Milky-Way from Figures 
3 and 4 of Cohen et al. (1980).}
\end{figure}

\clearpage

\begin{figure}
\figurenum{10}
\epsscale{1.0}
\plotone{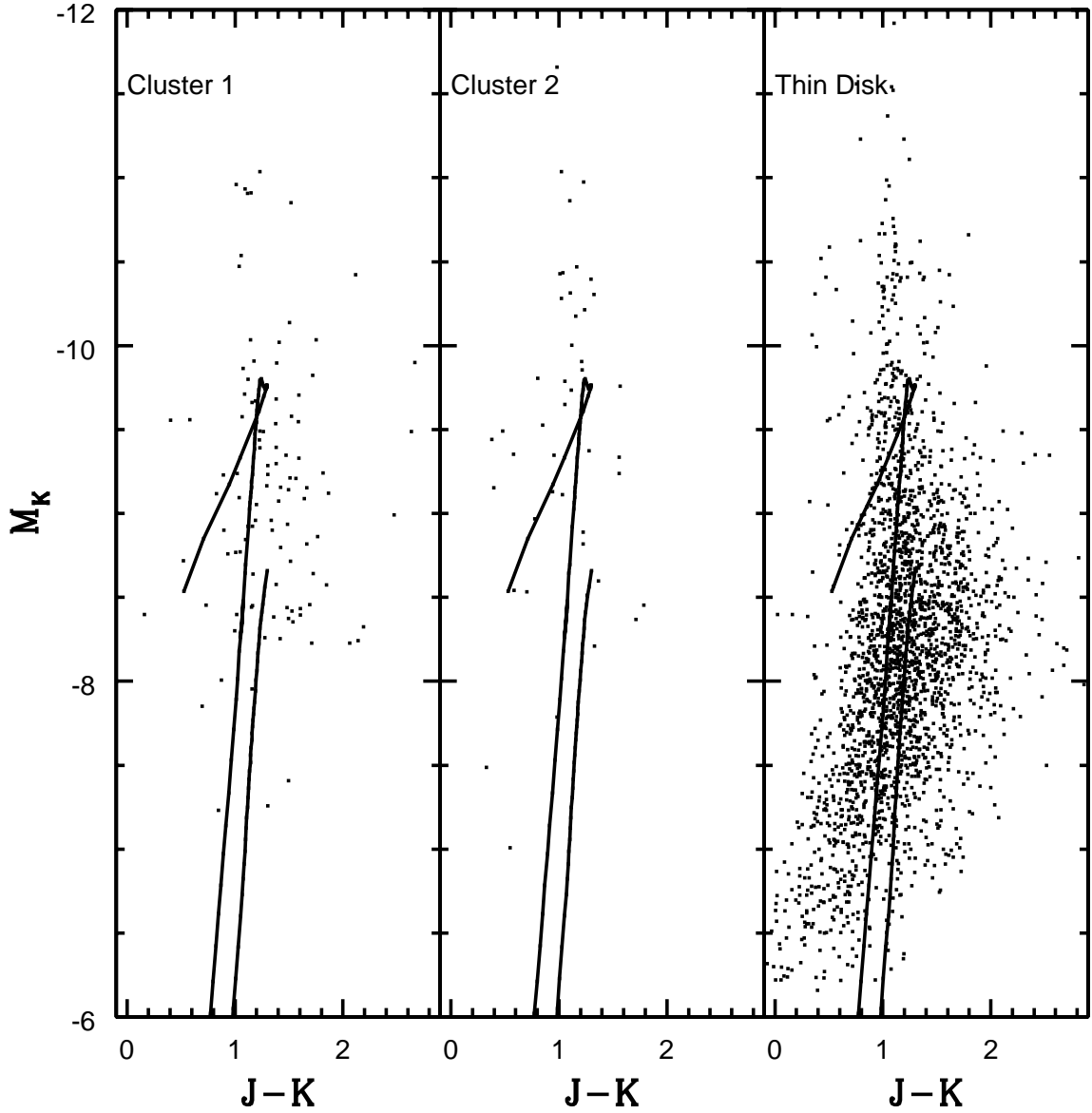}
\caption
{The $(M_K, J-K)$ CMD of Clusters 1 and 2 and the thin disk. A 
distance modulus of 26.5 has been assumed, based on the $i'$ 
brightness of the RGB-tip measured from the GMOS data. 
The solid lines are Z = 0.008 isochrones from Girardi et al. 
(2002) with log(t$_{yr}$) = 8.15 and 9.0.}
\end{figure}

\clearpage

\begin{figure}
\figurenum{11}
\epsscale{1.0}
\plotone{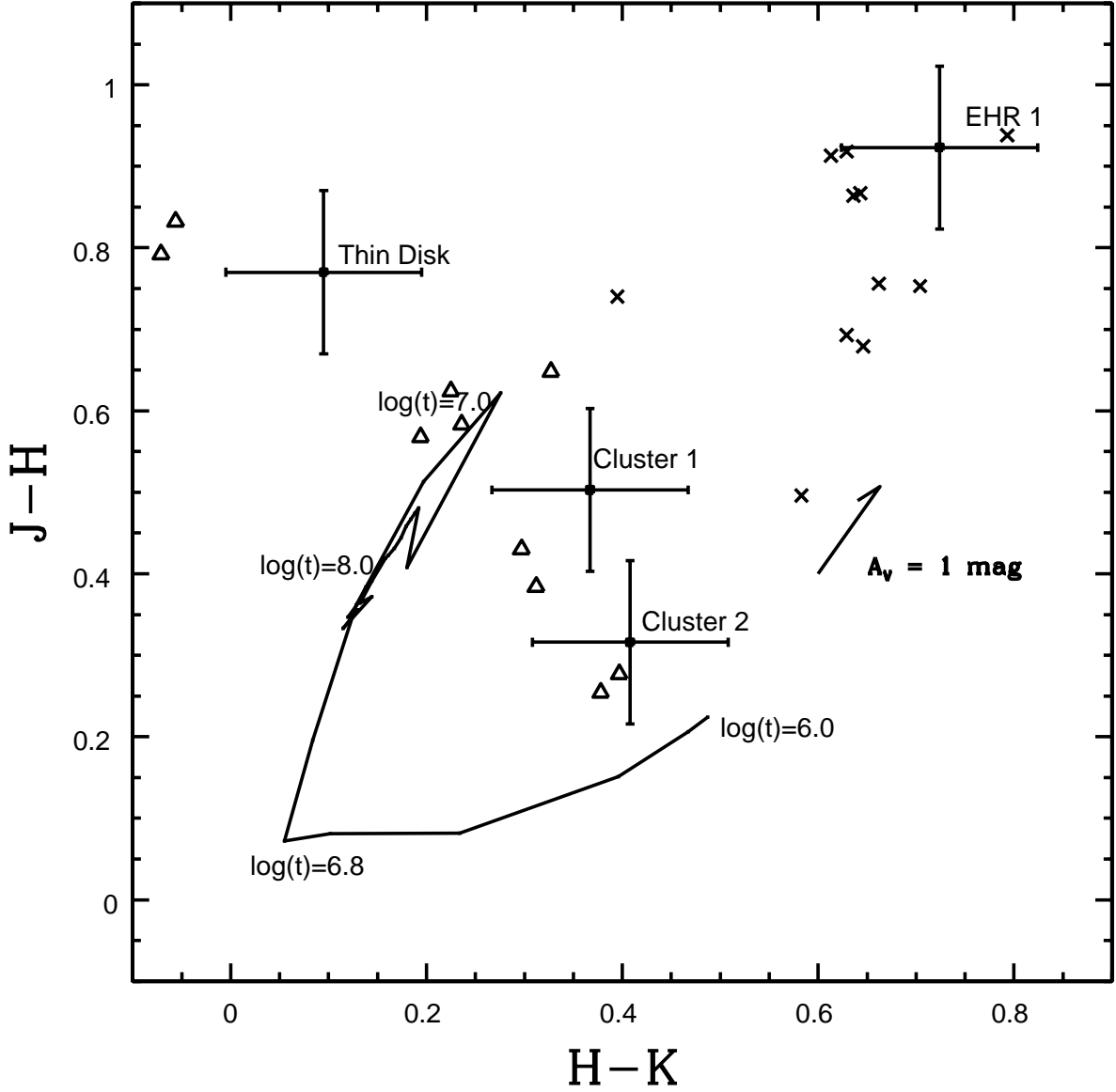}
\caption
{The $(J-H, H-K)$ TCD, showing the locations of Clusters 1 and 2, 
the thin disk near Cluster 1, and the extraplanar HII region 
EHR 1. The solid line shows the locus of Z=0.004 models from Leither et al. 
(1999) with $\alpha = 2.35$, and masses between 1 
and 100 M$_{\odot}$. Various ages along the sequence 
are indicated. The crosses are young clusters near 
the center of the star-forming galaxy NGC 3077 from Davidge 
(2004), while the open triangles are SB(s)m galaxies in the 2MASS Large Galaxy Atlas 
(Jarrett et al. 2003); the colors of the latter objects were computed from their total 
$J, H,$ and $K$ brightnesses. The reddening vector has a length that corresponds 
to A$_V = 1$ magnitude.}
\end{figure}

\clearpage

\begin{figure}
\figurenum{12}
\epsscale{1.0}
\plotone{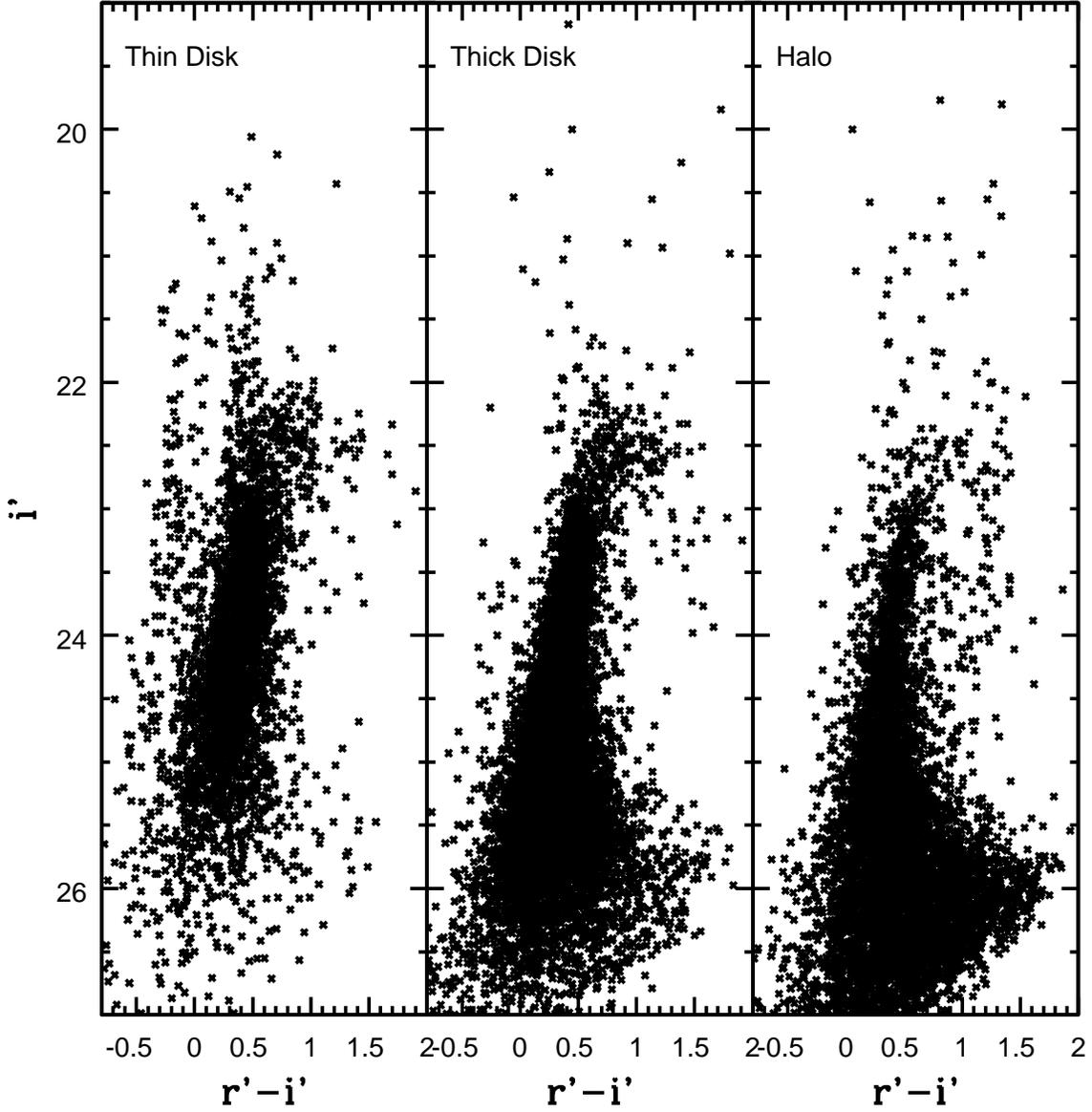}
\caption
{The $(i',r'-i')$ CMDs of the thin disk, thick disk, and halo 
portions of the GMOS NGC 55 field.}
\end{figure}

\clearpage

\begin{figure}
\figurenum{13}
\epsscale{1.0}
\plotone{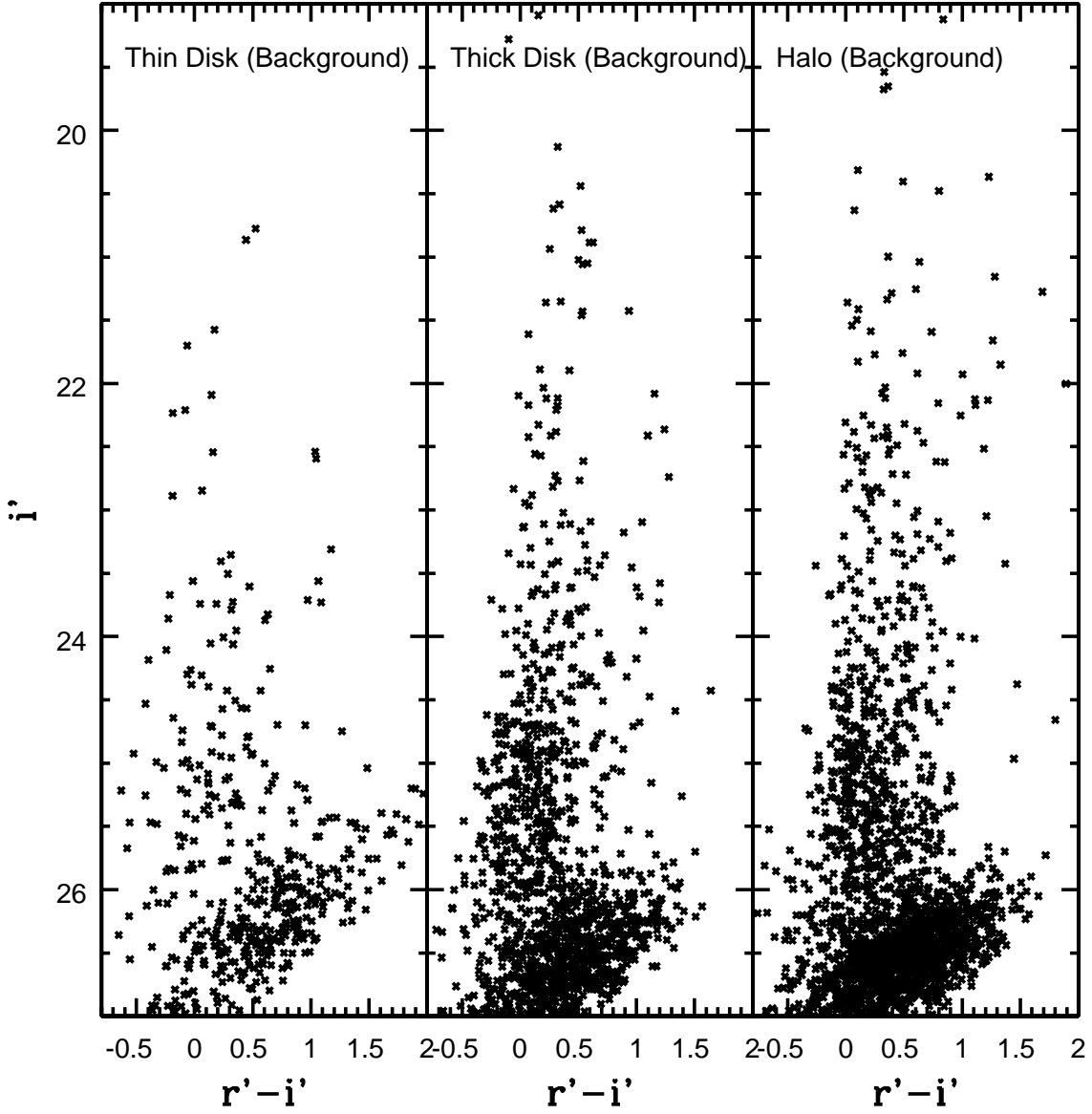}
\caption
{The $(i',r'-i')$ CMDs of areas in the GMOS control field that correspond to 
the thin disk, thick disk, and halo regions defined in NGC 55.}
\end{figure}

\clearpage

\begin{figure}
\figurenum{14}
\epsscale{1.0}
\plotone{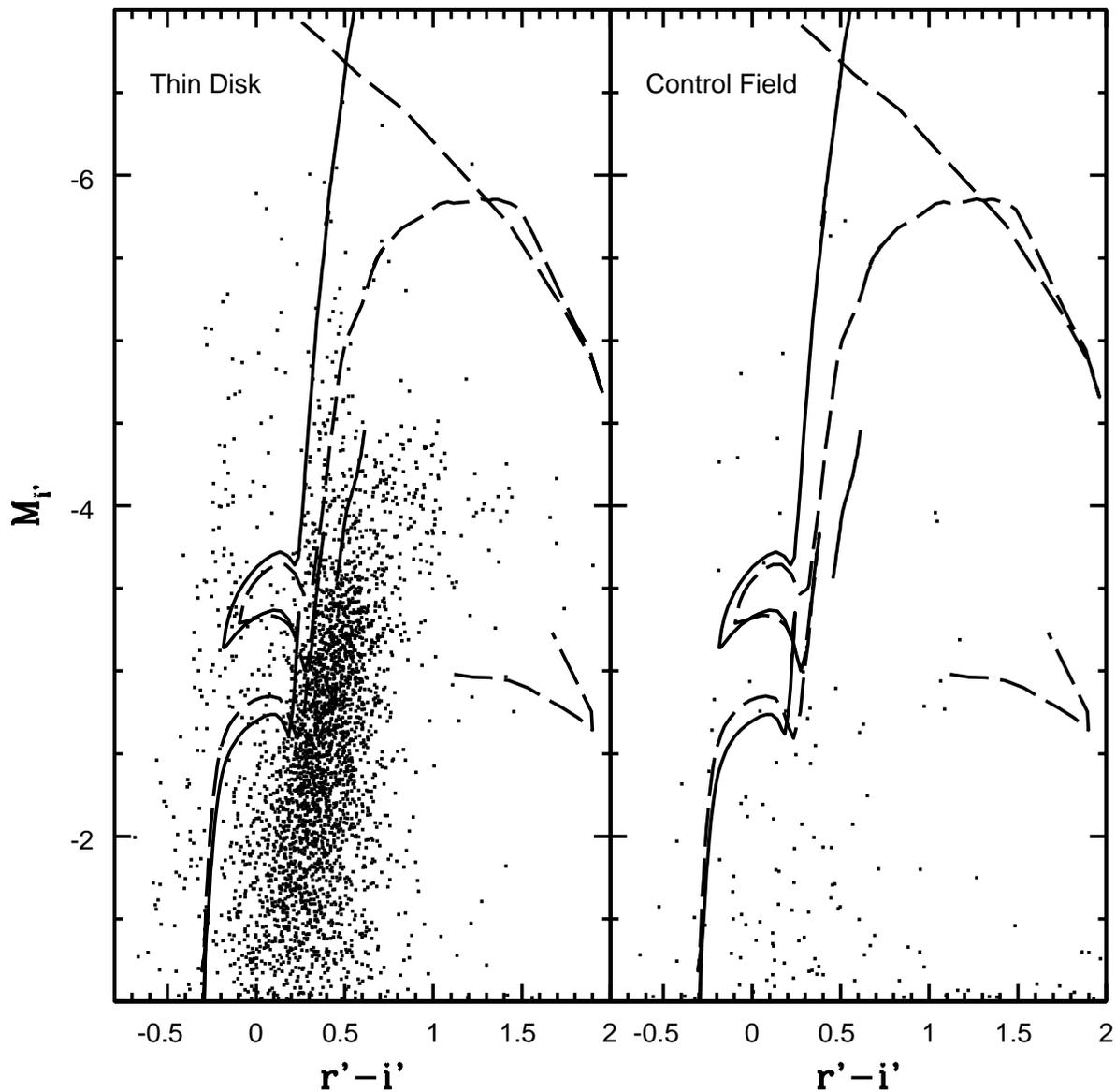}
\caption
{The $(M_{i'},r'-i')$ CMDs of the thin disk and the corresponding portion 
of the control field. A distance modulus of $\mu_0 = 26.5$ 
is assumed, based on the $i'$ brightness of the RGB-tip (\S 4.2). The 
solid lines are Z=0.008 isochrones with log(t) = 8.15 and 10.0 from Girardi 
et al. (2002), which were transformed into the SDSS filter system using 
relations from Fukugita et al. (1996). The dashed lines show isochrones 
with the same ages, but with Z=0.001. The log(t)=10 isochrones only 
include the AGB.}
\end{figure}

\clearpage

\begin{figure}
\figurenum{15}
\epsscale{1.0}
\plotone{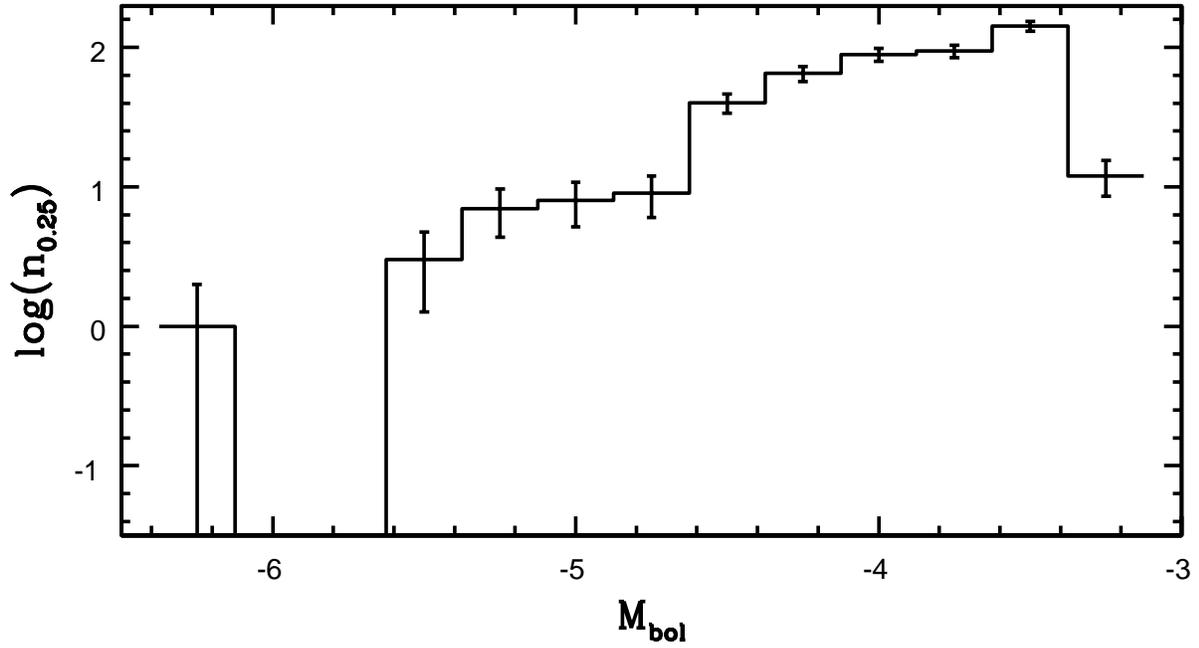}
\caption
{The bolometric LF of AGB stars in the thin 
disk, as computed from the GMOS data using the 
procedures described in the text. A distance modulus of 26.5 has been assumed, 
based on the $i'$ brightness of the RGB-tip (\S 4.2). N$_{0.25}$ is 
the number of stars per 0.25 mag interval in M$_{bol}$. 
Note the break in the LF near M$_{bol} = -4.6$. 
The error bars show counting statistics.}
\end{figure}

\clearpage

\begin{figure}
\figurenum{16}
\epsscale{1.0}
\plotone{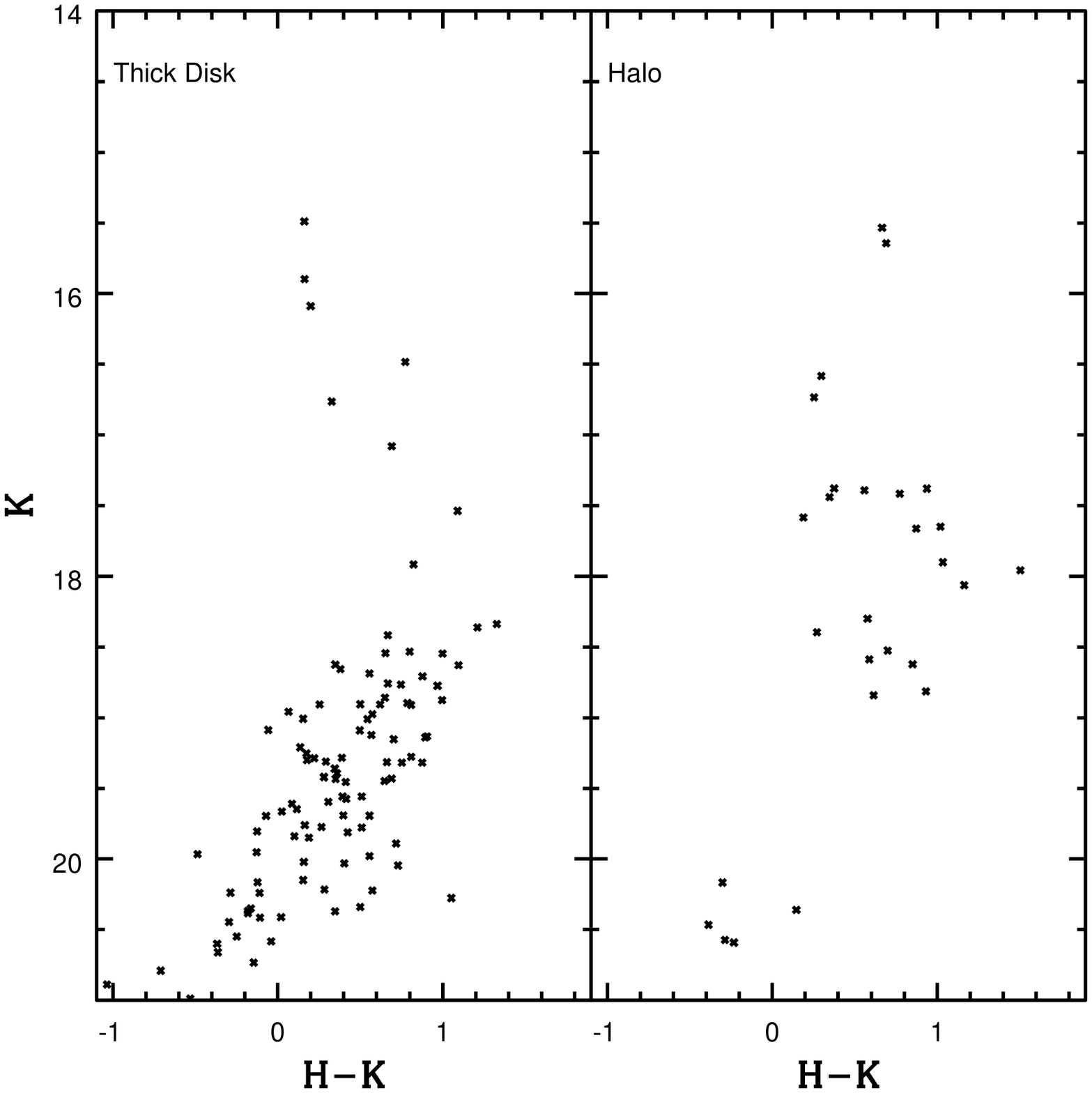}
\caption
{The $(K, H-K)$ CMDs of the thick disk and halo portions of the CFHTIR strip. 
It is shown in Figure 7 that the bright stars in these CMDs have SEDs that differ 
from those of evolved stars, and hence are likely not at the distance of NGC 55.}
\end{figure}

\clearpage

\begin{figure}
\figurenum{17}
\epsscale{1.0}
\plotone{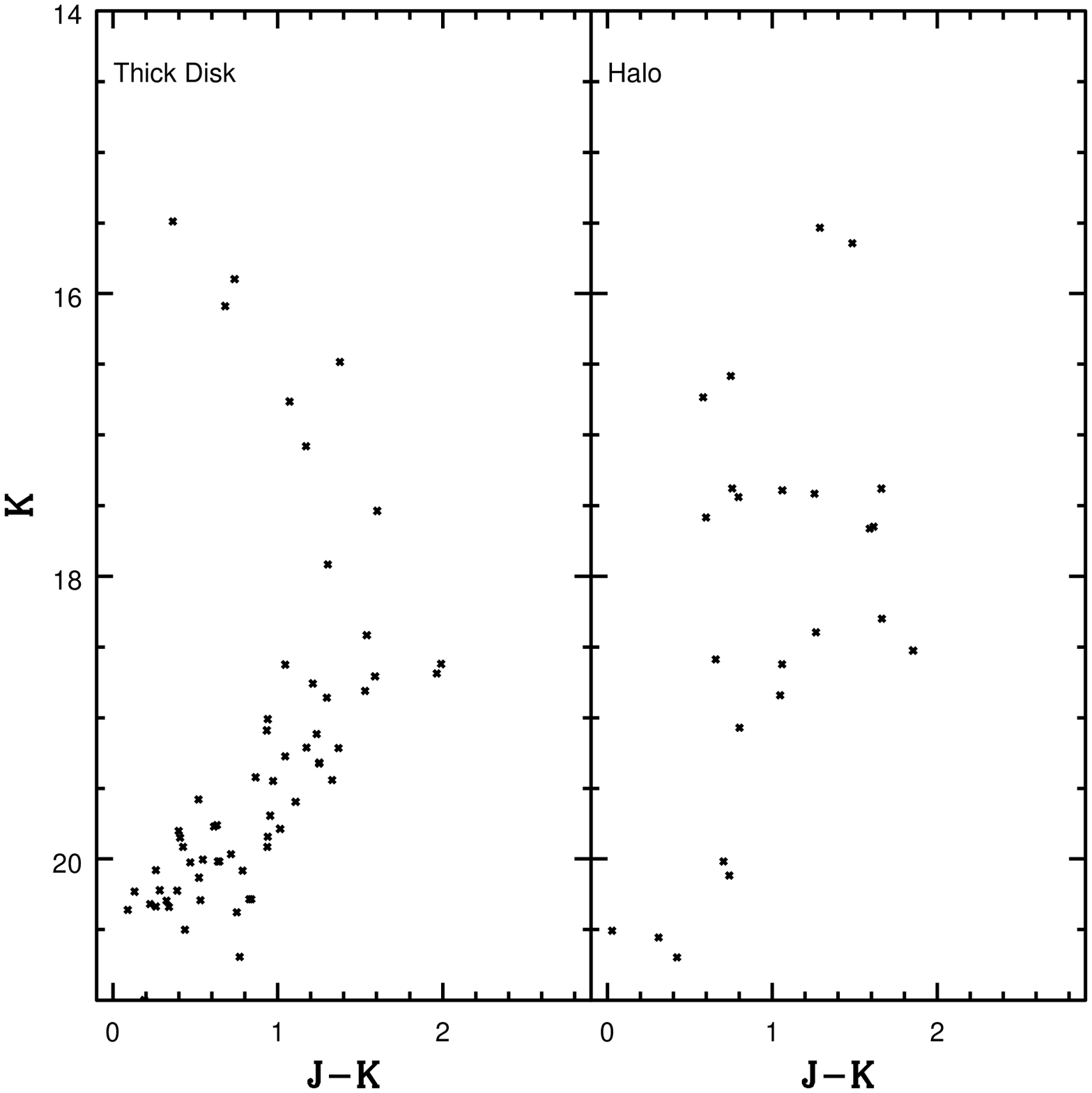}
\caption
{The $(K, J-K)$ CMDs of the thick disk and halo portions of the CFHTIR strip. 
It is shown in Figure 7 that the bright stars in these CMDs have SEDs that differ 
from those of evolved stars, and hence are likely not at the distance of NGC 55.}
\end{figure}

\clearpage

\begin{figure}
\figurenum{18}
\epsscale{1.0}
\plotone{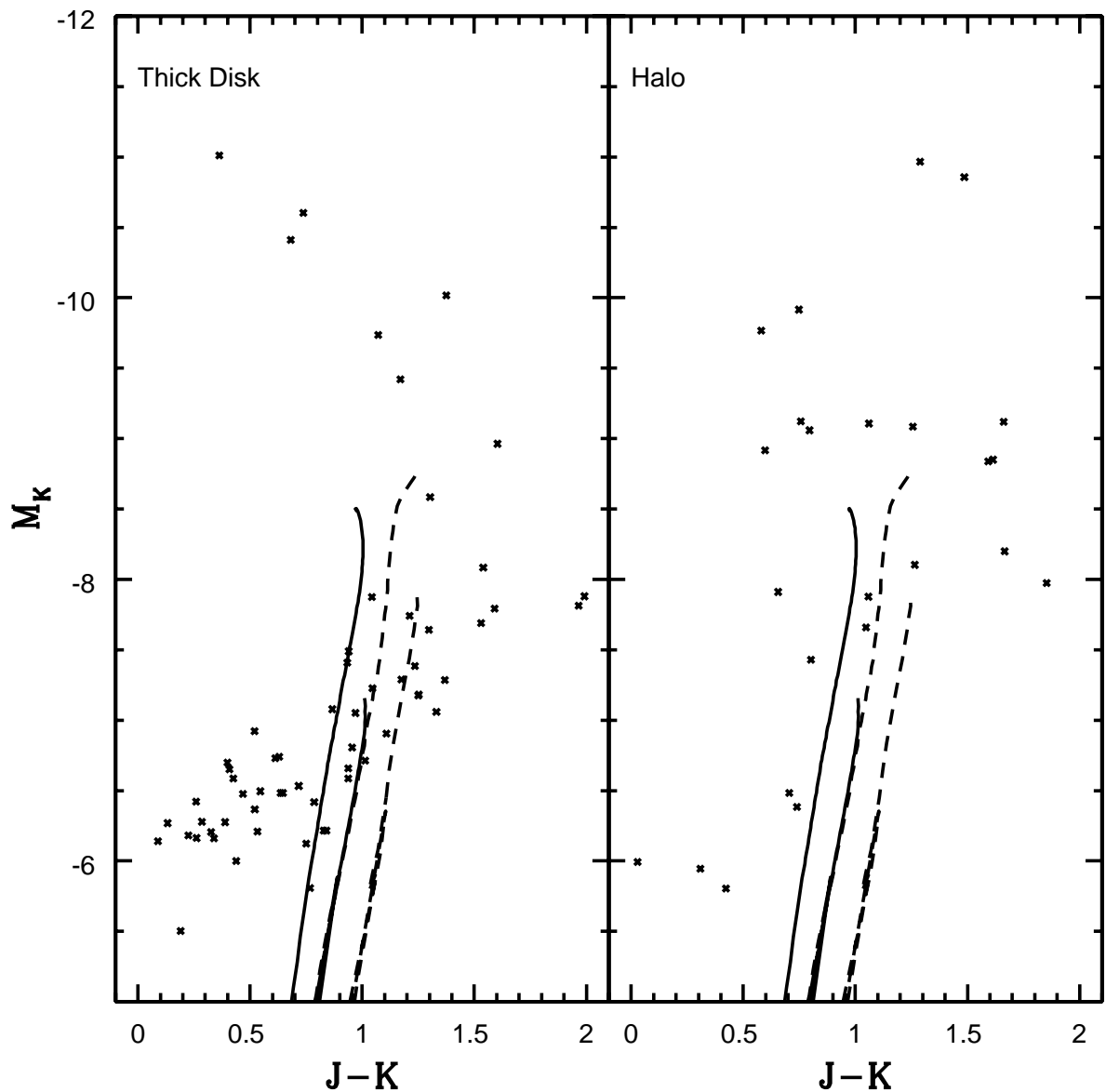}
\caption
{The $(M_K, J-K)$ CMD of the thick disk and halo portion of the CFHTIR strip. A 
distance modulus of 26.5 has been assumed, based on the $i'$ brightness of 
the RGB-tip (Section 4.2). The solid lines are Z = 0.001 isochrones 
from Girardi et al. (2002) with log(t$_{yr}$) = 9.0 and 10.0, while 
the dashed lines are Z = 0.004 isochrones with the same ages.}
\end{figure}

\clearpage

\begin{figure}
\figurenum{19}
\epsscale{1.0}
\plotone{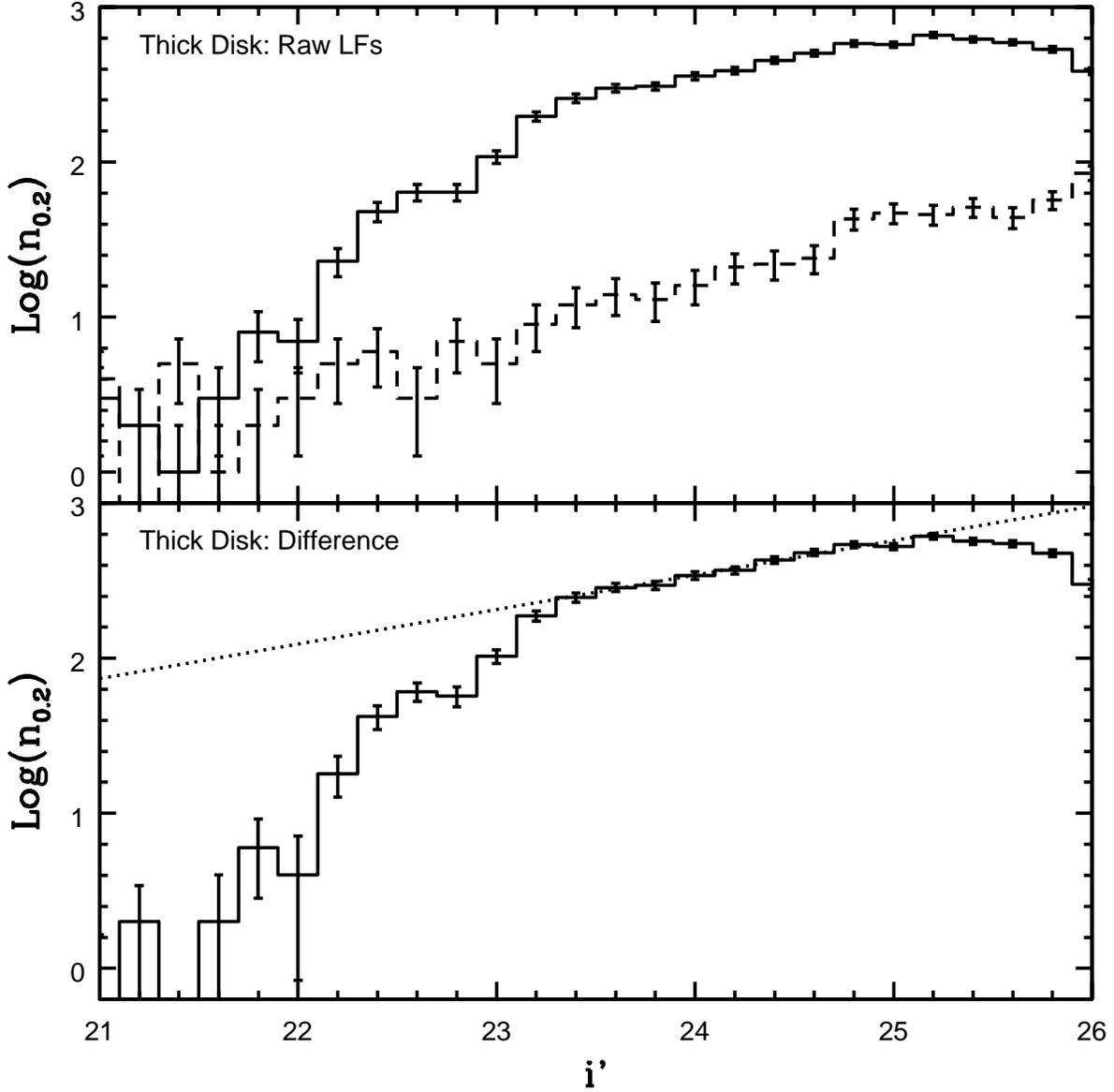}
\caption
{The top panel shows the $i'$ LF of sources in the thick disk portion
of the NGC 55 GMOS field (solid line) and the corresponding area of the 
control field (dashed line). n$_{0.2}$ is the number of stars per 0.2 magnitude 
interval in $i'$ with $r'-i'$ between --0.5 and 1.5. The error bars show 
uncertainties due to Poisson statistics. The lower panel shows the NGC 55 LF 
after subtracting the control field LF. The dotted line is a power-law with an 
exponent $x = 0.22 \pm 0.01$, which was computed from the 
entries in the lower panel with $i'$ between 23.5 and 25.0. Note that the 
LF departs significantly from the fitted power-law near $i' = 23.1$, and this is 
adopted as the brightness of the RGB-tip.}
\end{figure}

\clearpage

\begin{figure}
\figurenum{20}
\epsscale{1.0}
\plotone{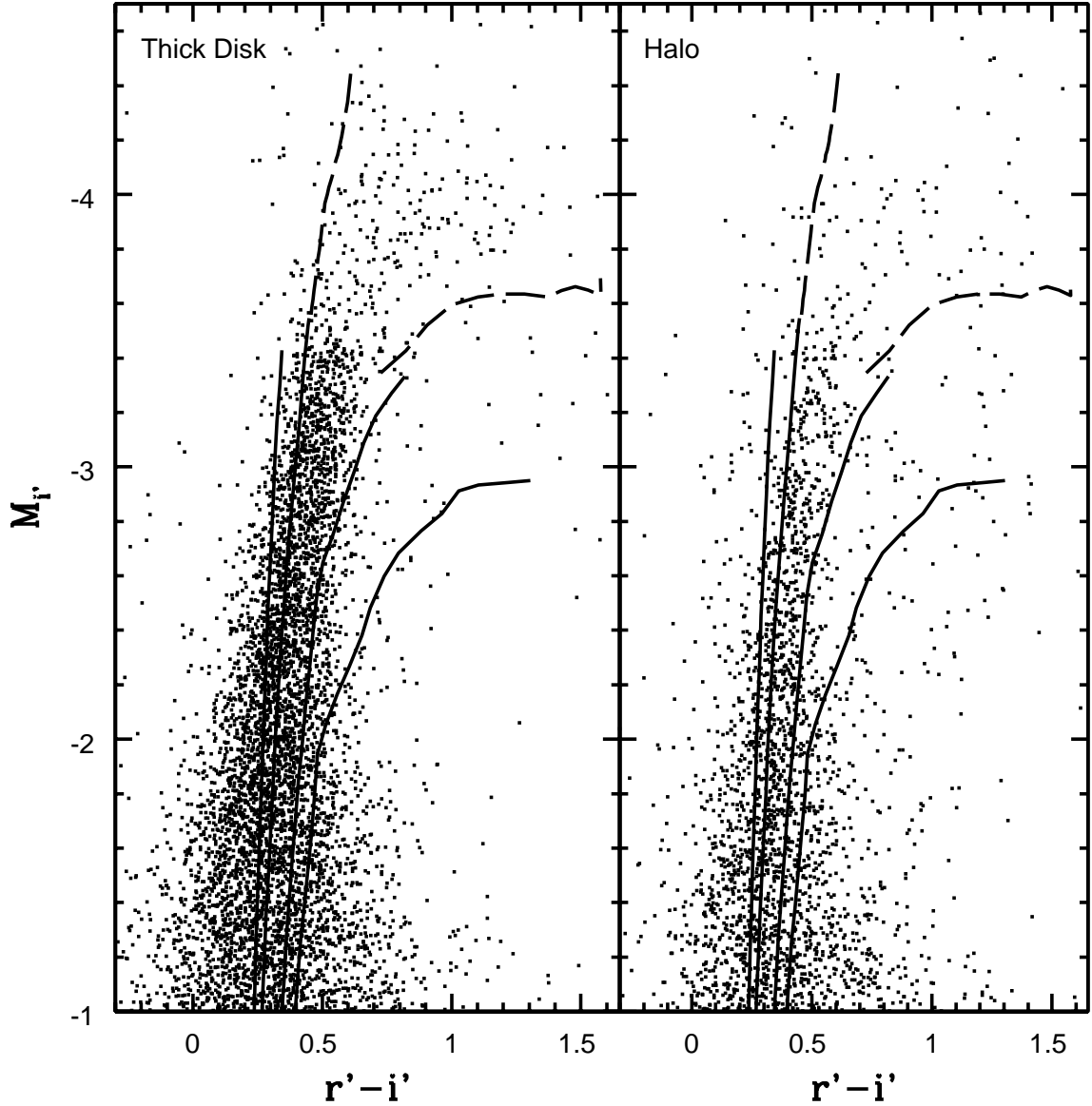}
\caption
{The $(M_{i'}, r'-i')$ CMD of the thick disk and halo of NGC 55, assuming a 
distance modulus of 26.5 based on the $i'$ brightness of the RGB-tip. The 
solid lines are RGB isochrones from Girardi et al. (2002) with log(t$_{yr}$)=10 and 
Z = 0.0001, 0.001, 0.004 and 0.008. These isochrones were transformed into the 
SDSS filter system using relations from Fukugita et al. (1996). The 
dashed lines show the AGB extensions of the Z=0.001 and Z=0.004 
isochrones. Note that the majority of RGB stars 
have a metallicity between Z=0.001 and 0.004, and that the brightest AGB 
stars have an age log(t$_{yr}) = 10 \pm 0.1$ dex (see text).}
\end{figure}

\end{document}